**Self-organized Natural Roads for Predicting Traffic Flow: A Sensitivity Study**

Bin Jiang, Sijian Zhao and Junjun Yin

Department of Land Surveying and Geo-informatics
The Hong Kong Polytechnic University, Hung Hom, Kowloon, Hong Kong
Email: bin.jiang@polyu.edu.hk

**Abstract**
In this paper, we extended road-based topological analysis to both nationwide and urban road networks, and concentrated on a sensitivity study with respect to the formation of self-organized natural roads based on the Gestalt principle of good continuity. Both Annual Average Daily Traffic (AADT) and Global Positioning System (GPS) data were used to correlate with a series of ranking metrics including five centrality-based metrics and two PageRank metrics. It was found that there exists a tipping point from segment-based to road-based network topology in terms of correlation between ranking metrics and their traffic. To our big surprise, (1) this correlation is significantly improved if a selfish rather than utopian strategy is adopted in forming the self-organized natural roads, and (2) point-based metrics assigned by summation into individual roads tend to have a much better correlation with traffic flow than line-based metrics. These counter-intuitive surprising findings constitute emergent properties of self-organized natural roads, which are intelligent enough for predicting traffic flow, thus shedding substantial insights into the understanding of road networks and their traffic from the perspective of complex networks.

## 1. Introduction

Natural roads are joined road segments based on the Gestalt principle of good continuity, and they are self-organized in nature. Let us assume that every segment at each end chooses one most suitable neighboring segment with a smallest deflection angle to join together; and this process (refer to Figure 16 later in the text for an illustration) goes on until the deflection angle is greater than a preset threshold (e.g. degree 45). This process resembles Bak's sandpile (Bak, Tang and Wiesenfield 1987; Bak 1996) in which sand is added continuously to generate different sizes of avalanche. The size of avalanches exhibits a universal regularity of power law distribution, the same behavior demonstrated by the connectivity and length of natural roads (c.f. a later section on more discussions). There is also no typical size of natural roads. More than the avalanches, natural roads demonstrate a sort of collective intelligence (Surowiecki 2004) that is able to predict traffic flow.

Self-organized natural roads, or strokes in terms of Thomson (2003), differ from named roads that are identified by unique names (Jiang and Claramunt 2004). Named roads are more difficult to implement than natural roads because of the incomplete nature of road databases, in which some segments may have missing or wrong names. On the other hand, the formation of natural roads is significantly biased by join principles and deflection angle threshold in the join process. In other words, there is a sensitivity issue involved in the formation of natural roads.

How each segment determines to join with one of its neighboring segments follows a self-organized process based on three different join principles. The first is called every-best-fit, and it works like this. Every pair of segments at a junction point have to negotiate with each other to have best fit (i.e., the one with a smallest deflection angle), in terms of which one joins which one. This principle is rather utopian or communism in nature, and seems the best strategy. The second is called self-best-fit. Instead of *every*, each segment only considers it*self* to find a best fit, and does not care about others in the process. Thus it is rather selfish or capitalism in nature. Or it can be deemed a natural selection. Similar to this selfish principle, there is another one called self-fit. Obviously each segment tries to choose arbitrarily one fit, i.e., the one with a deflection angle less than a preset threshold, to join, but not necessarily to be the best fit. In comparison (c.f. Figure 1 for an illustration, and Appendix B for algorithms), the first principle always leads to a unique set of natural roads, while the other two principles would generate enormous sets of natural roads, depending on the search order of the segments. Figures 1c and 1d are just one of many possible sets for each principle. It is one of the sensitivity issues we intend to study in the paper.

This paper is intended to investigate the join principles and deflection angle threshold with respect to the formation of natural roads, and their correlation to Annual Average Daily Traffic (AADT) and GPS data (both referred to as traffic flow or flow in what follows). We found that there exists a tipping point from segment-based to road-based network topology in terms of correlation between ranking metrics and their traffic flow (or metric-flow correlation). To our big surprise, (1) the correlation based on the principles of self-best-fit and self-fit is much better than that based on the principle of every-best-fit, and (2) point-based metrics (in particular for local and global integrations, see Appendix C) assigned by summation into individual roads tend to have a much better correlation with traffic flow than line-based metrics. These counter-intuitive surprising findings provide telling evidence that self-organized natural roads are an emergence developed from individual road segments, and are intelligent enough for predicting traffic flow, thus shedding substantial insights into the understanding of road networks and their traffic from the perspective of complex networks.

The remainder of this paper is structured as follows. Section 2 introduces geometric and topological representations of road networks using a notional road network, in particular different transformations from geometric to topological representation using the three join principles. We brief data sources and essential processing of road networks as well as observed traffic in section 3. Section 4 illustrates our experiments and findings about various sensitivity issues. Section 5 speculates in detail on the emergent properties of natural roads and their implications. Finally section 6 concludes the paper with a summary and future work.

## 2. Geometric versus topological representations of road networks

Although road networks can be abstracted as graphs, represented by a Point-Point Distance Matrix (PPDM, c.f. A-1 for an example), we still refer to them as a geometric representation (the network in gray in Figure 1a). This is based on the facts that (1) the junction points have precise geometric coordinates referenced to the earth, and (2) the distances between the pairs of points are a major concern for the representation. The points are defined in a Euclidian space, and the distances are taken by the matrix as its elements. Even though the distance matrix can be further abstracted topologically into a Point-Point Connectivity Matrix (PPCM, c.f. A-2 or the connectivity graph in Figure 1a), it still cannot be regarded as a true topology because of a lack of an interesting structure or pattern. We can remark that the networks or graphs have a very boring connectivity structure, because of the lack of variation in connectivity for individual points. The same observation can be made in reality where most junctions have a degree of 4, and a very few have a degree less or greater than 4. However, things would be rather different if we take a truly topological view (c.f. Figure 4 also).

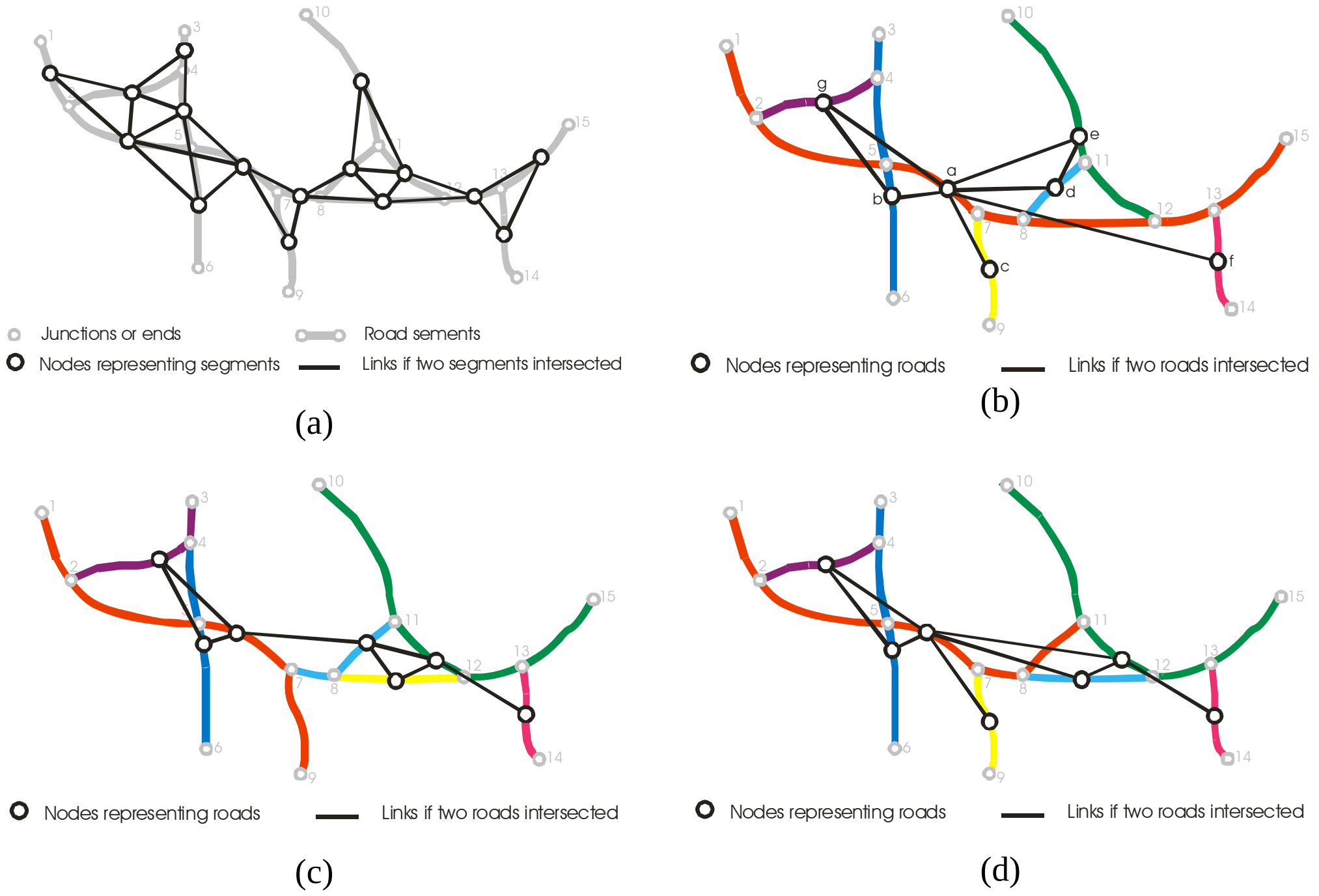

Figure 1: (color online) A notional road network and its connectivity graphs: (a) segment-based connectivity graph, and road-based connectivity graphs with respect to different join principles of (b) every-best-fit, (c) self-best-fit, and (d) self-fit

This topological view takes a higher level (macro-scale) of abstraction, which considers adjacent relationships of individual roads, i.e., a sort of road-road intersection. Seven intersected roads can be transformed into a Line-Line Adjacency Matrix (LLAM, c.f. A-3), which is equivalent to the connectivity graphs shown in Figure 1b, Figure 1c and Figure 1d with respect to the join principles of every-best-fit, self-best-fit and self-fit. The matrix contains all information about the adjacency: its elements are set to 1 if the corresponding roads are intersected and 0 otherwise. The graph based on LLAM contains no geometric information but the binary relation (1/0), i.e., no coordinates or distances attached to the nodes and links. However, the graph demonstrates some interesting connectivity structure, e.g., the distribution of connectivity of the nodes is skewed significantly. It sets a clear difference from the underlying road networks.

Another topological representation in terms of point and point relationship can be developed. The point-point relationship is set up based on whether or not a pair of points share a road in common, i.e. 1 if yes and 0 otherwise in the corresponding Point-Point Adjacency Matrix (PPAM, c.f. A-4). This alternative graph also demonstrates an interesting structure in terms of variation of connectivity. It is important to note the relationship of the two topological representations. They are closely related, and can be easily derived through the operation of multiplication of Line-Point Incidence Matrix (LPIM, c.f. A-5) and its transpose.

The LLAM-based representation is well developed and applied in space syntax community (Hillier and Hanson 1984), but it is far less so for the PPAM based representation. For sake of simplicity and intuition, we will respectively call them line-based and point-based approaches (Jiang and Claramunt 2002). Alternatively, they are named as primal and dual representations (Batty 2004; Jiang and Liu 2007). They are a powerful tool for obtaining structure and patterns, thus an important analytical model for predicting traffic flow (e.g., Jiang and Liu 2007). However, the dual relationship has yet to be applied, and a sensitivity study about the prediction deserves further investigation as shown later in this paper.

## 3. Data sources and processing

A main dataset for the study was obtained from the Swedish Road Administration (Vägverket), and it contains both road networks and AADT assigned to each individual road segment. It should be noted that it is a massive dataset, involving in total ~ 45 000 segments, ~ 100 000 kilometers in length, across Sweden (Figure 2a). The entire road network is divided into seven regions (Figure 2b). We keep the seven networks for separate investigations for consistent checking of our findings, and in the mean time merge them together as an entire network for some experiments. Before the experiments, isolated segments were removed to ensure all roads are interconnected. It is important to note that the percentage of isolated segments is very low (< 0.5%), except for the region of Stockholm, which is a bit higher. This is probably a partial reason that it shows some special behavior compared with other regions. As for the Gävle urban street network (Figure 2c), it consists of ~3 400 segments, and traffic flow is obtained from GPS log files, by one taxi company (Taxi 2007), recording locations of 50 taxicabs every 10 seconds. The GPS dataset has been preprocessed to ensure the recoded locations are truly trajectories. We have in total seven days (October1-7, 2007) of such data for consistent checking, the same data used in Jiang (2006). Before any further experiments, we make sure the networks are truly road segment based. In case they are not, we join separate parts together to be one segment between two junctions, in order to form the kind of road network shown in Figure 1. In the course of this process, traffic flows of the separate parts are averaged to the segment.

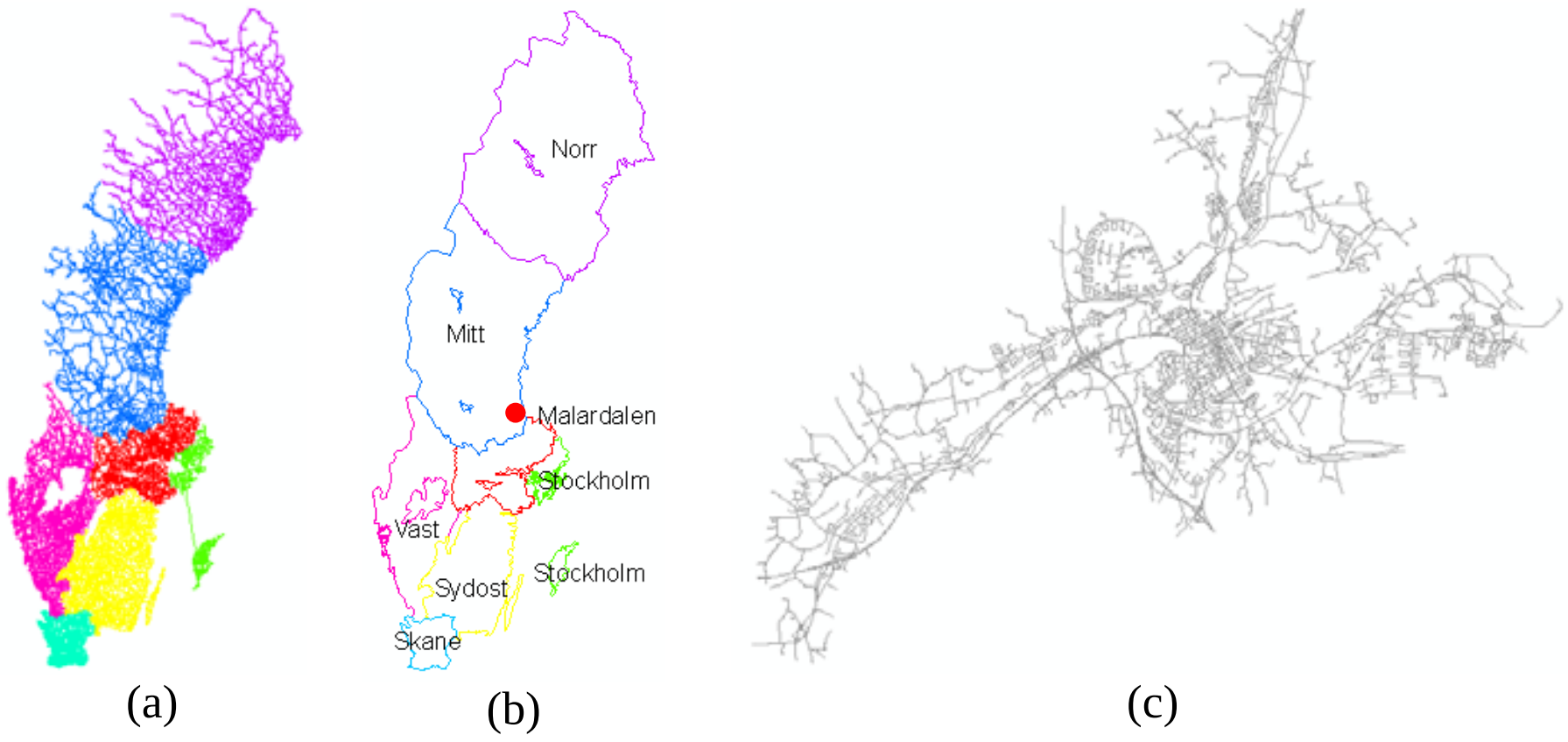

Figure 2: (color online) nationwide road networks in Sweden (a) divided into seven regions (b) and Gävle urban street network (c) (NOTE: the red spot in b is the location of the city Gävle)

## 4. Experiments and findings

### 4.1 Overall statistics on segments versus roads

We first examine the sensitivity of deflection angle threshold in forming natural streets. We chose every fifth degree as an interval between degree 0 and 90 to examine how many roads generated from individual segments with respect to the series of threshold angles. As plotted in Figure 3, the number of roads drops from degree 0 to degree 5 dramatically, and continuously yet slowly till degree 30. And, the number tends to become rather stable from degree 30 onwards. This observation is valid for both nationwide and urban road networks. Intuitively, degree 45 appears to be an ideal threshold angle that helps generate natural roads with a good continuity. The number of roads becomes stable around ~15 000 for the nationwide network, and around ~ 1 100 for the urban street network. In this respect, both the nationwide road network and urban street network shows a significant similarity.

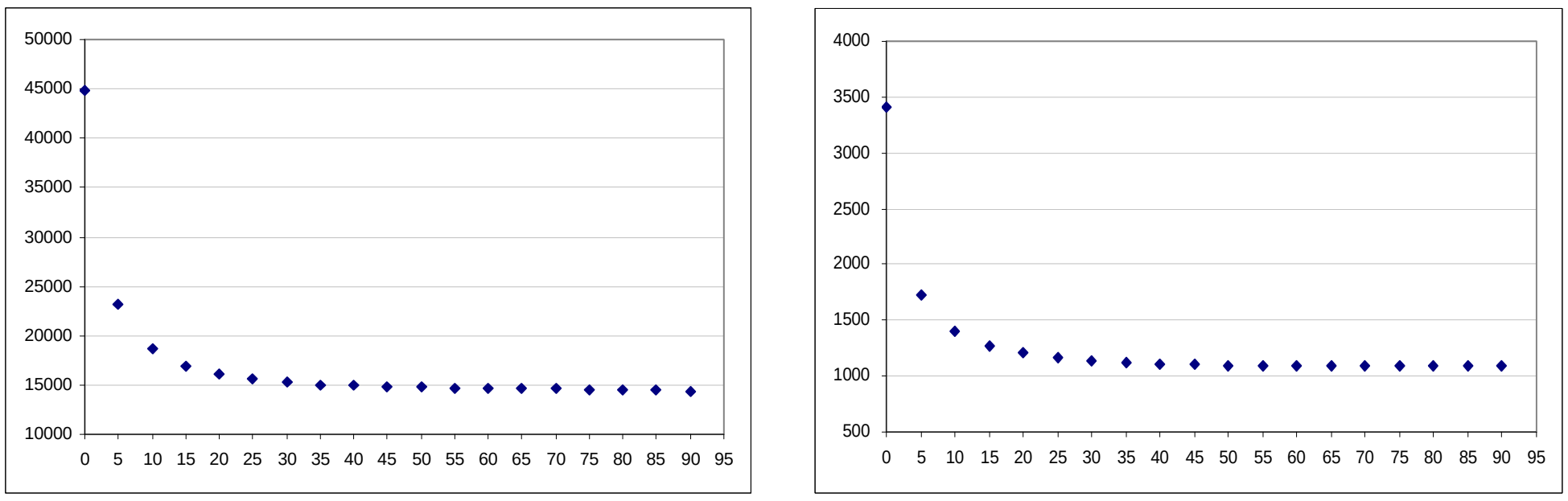

(a) (b)

Figure 3: The number of natural roads drops as the threshold angle rises: the case of the entire nationwide road network (a) and Gävle urban street network (b)

It is important to note the fact that the natural roads generated by the threshold angle 0 are identical to the segments. There are a couple of exceptions, where two adjacent segments have no angle change at all. The distribution of segment and road connectivity (when the threshold angle is set to 45) is very different: the former is a normal like distribution, while the latter is a power law distribution (Figure 4). We can remark that over 60% of segments have a connectivity of 4 (i.e., a typical connectivity), and maximum connectivity is not more than 11. On the other hand, maximum road connectivity is over 220, over 80% of roads have a connectivity less than 4, and there is no typical connectivity for natural roads. This diversity of road length and connectivity has been illustrated in a previous study, with a big sample of American cities and expressed by the 80/20 principle (Jiang 2007). Other researchers (e.g. Gastner and Newman 2006; Porta, Crucitti and Latora 2006) have also studied spatial networks from the perspective of complex networks with some interesting findings.

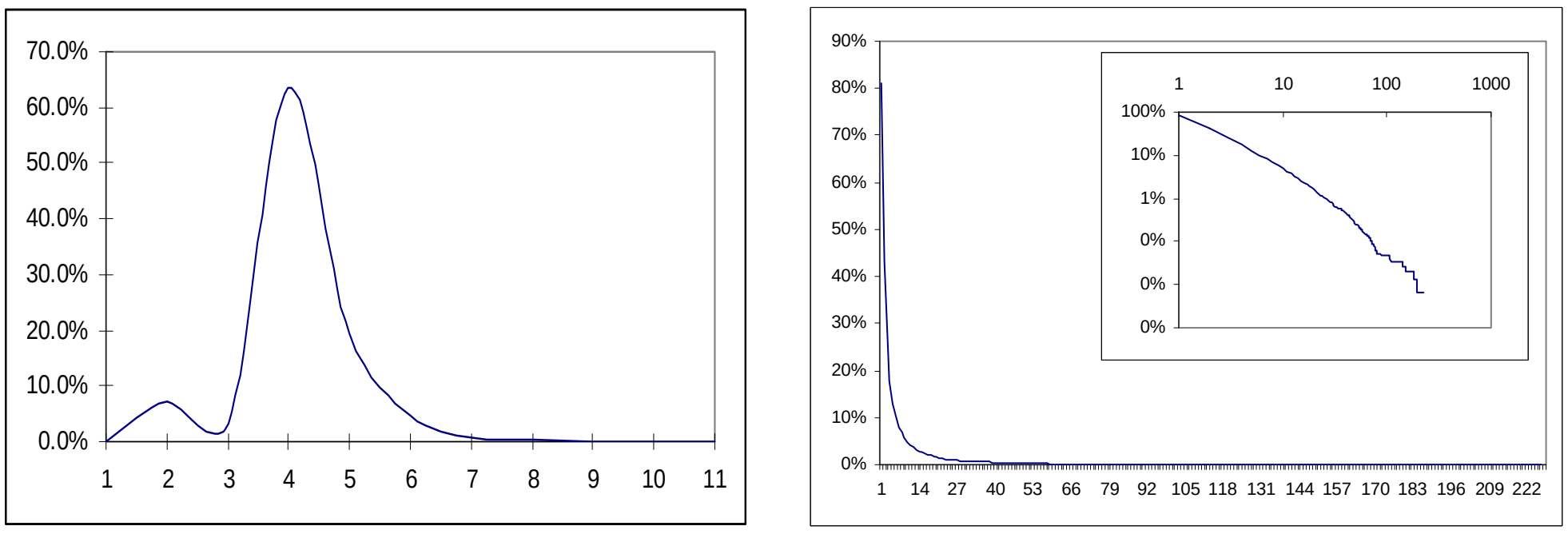

(a) (b)

Figure 4: Distribution of (a) segment connectivity and (b) road connectivity, whose log-log plot shows a straight line (the inset)

### 4.2 Findings based on the line-based approach

In what follows, we will demonstrate how ranking metrics (c.f. Appendix C) correlate to traffic flow, and how the metric-flow correlation alters with respect to the threshold angle and join principles. Before that, we examine whether or not the distribution of the metrics and traffic flow shows a universal regularity of power law with respect to the three join principles. From the plots in Figure 5, we can observe that except local and global integrations, all other metrics, as well as traffic flow (threshold angle is set to 45), exhibit a power law distribution. Clearly, natural roads generated by the principle of self-best-fit have a more striking power law than the other two. Overall, all the power law distributions are very similar, but there is a striking similarity between that of PageRank and connectivity (or control) metrics, and between flow and betweenness (or weighted PageRank) metrics. This is not particularly surprising, since (1) for an undirected graph PageRank scores reflect connectivity, and (2) the definition of betweenness metric is based on the concept of flow.

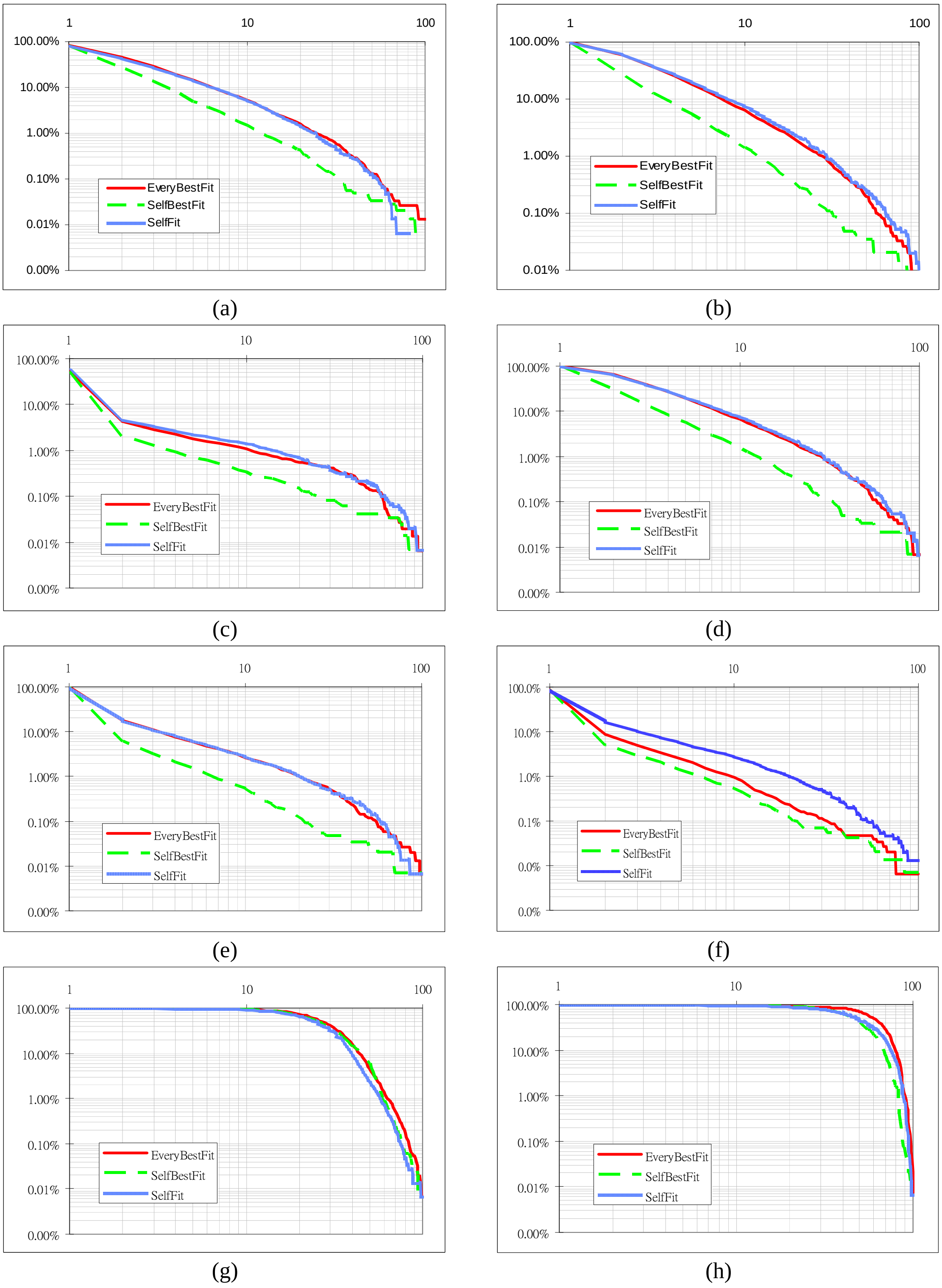

Figure 5: (color online) Log-log plots of (a) connectivity, (b) control, (c) betweenness, (d) PageRank (d = 0.20), (e) weighted PageRank (d = 0.20), (f) flow (threshold angle = 45), (g) local integration and (h) global integration

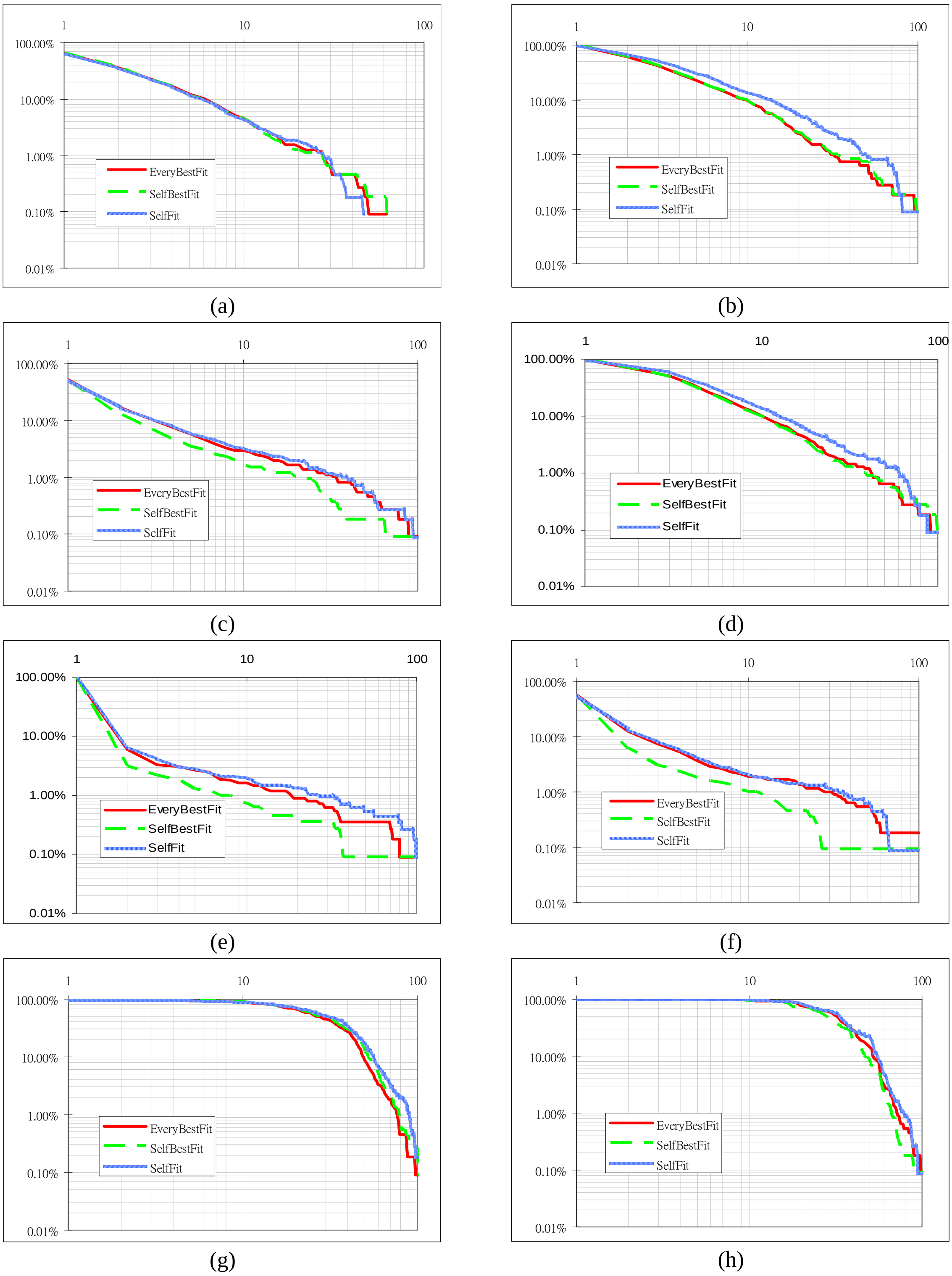

Figure 6: (color online) Log-log plots of (a) connectivity, (b) control, (c) betweenness, (d) PageRank (d = 0.95), (e) weighted PageRank (d = 0.95), (f) flow (threshold angle = 45), (g) local integration and (h) global integration

To fully examine the metric-flow correlation, we applied them to the nationwide road networks. Figures 7, 8 and 9 demonstrate one set of results with respect to the principles of every-best-fit, self-best-fit and self-fit for the region of Sydost. We can observe that in all three cases segments (threshold angle = 0) have no metric-flow correlation at all, with R square values being 0. However, this correlation rises gradually with the increase of the threshold angle until degree 15, and then becomes stable for a while. It appears that degree 15 is a kind of tipping point where R square values reach a maximum and become stable until degree 90. Of the seven metrics, weighted PageRank, PageRank, connectivity, and control (with a decreasing order) are the best ones in terms of metric-flow correlation. The poorest are local and global integrations, while the betweenness metric is somewhere between the best and poorest. Cross checking Figures 7, 8 and 9, self-best-fit (Figure 8) is the best option (R square over 0.75). We can also note that the damping factor d around 0.20 seems the best choice for the nationwide networks. It is important to note that in theory connectivity is equivalent to PageRank for an undirected graph. However, due to the different damping factor d values, the actual metric-flow correlations may not be identical. This is clearly reflected in the plots. The observations are pretty consistent among the seven regions of the nationwide road network. Thus, only one region is used for illustration purpose.

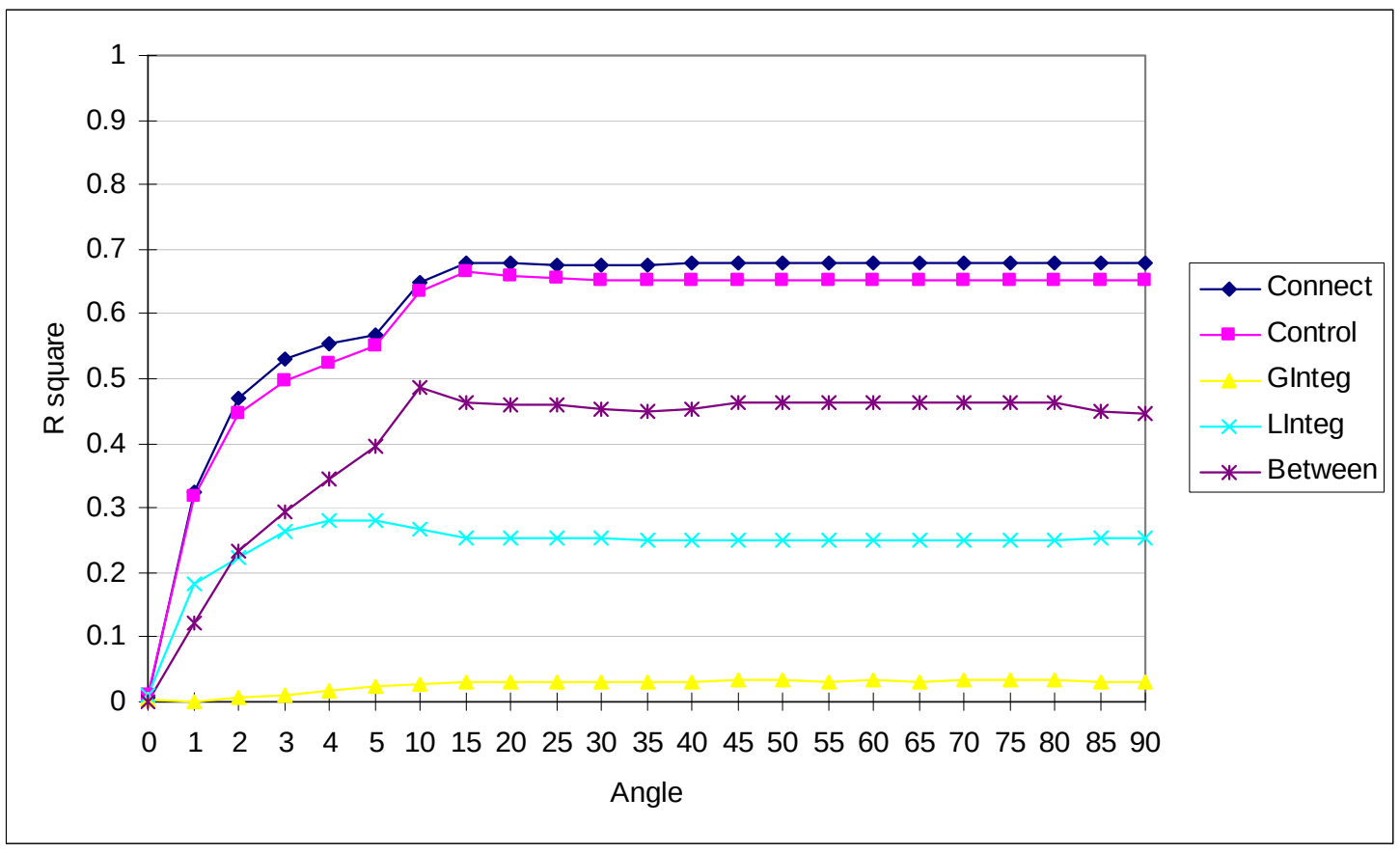

(a)

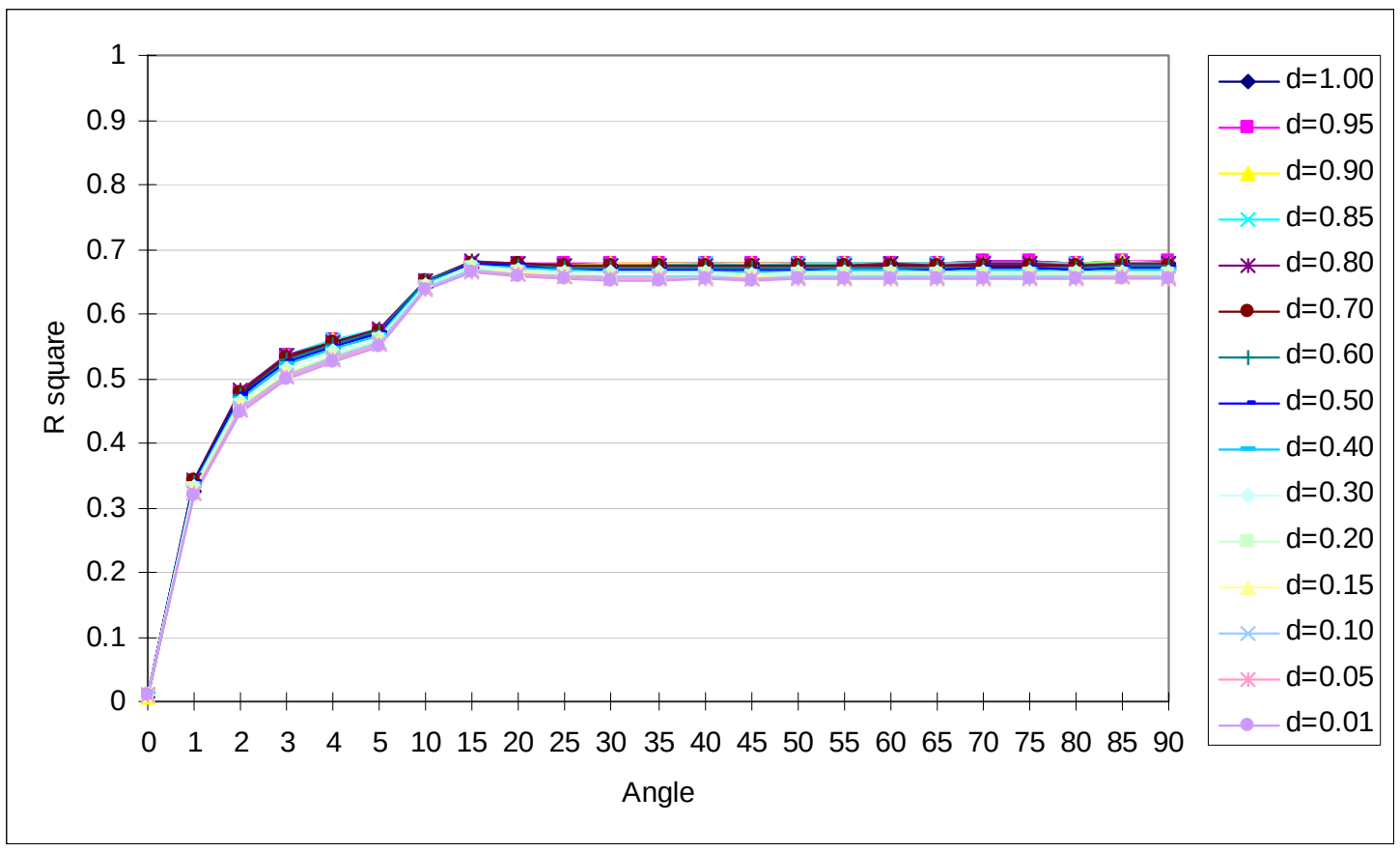

(b)

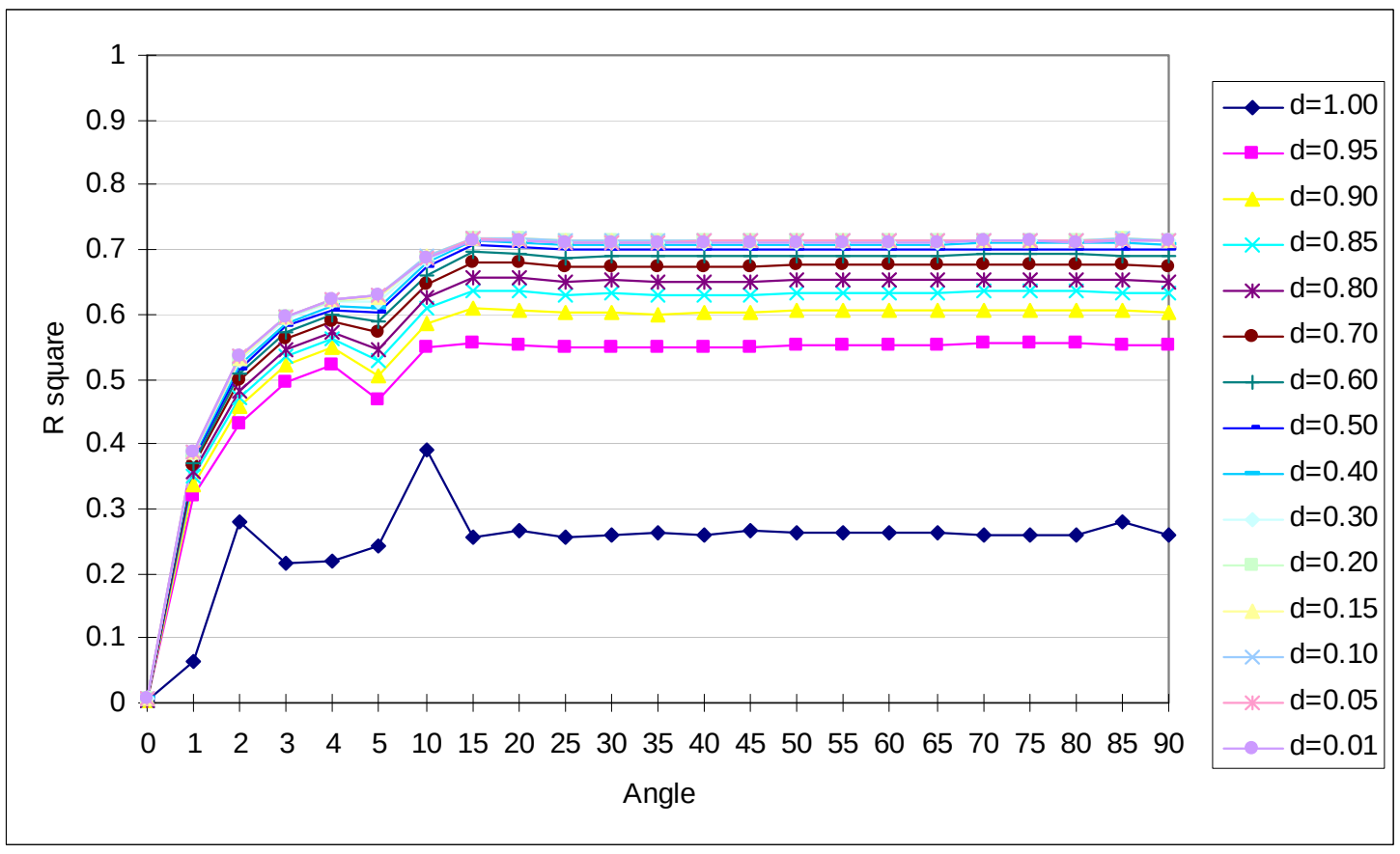

(c)

Figure 7: (color online) Correlation coefficient (R square) between traffic flow and (a) five centrality-based metrics, (b) PageRank and (c) weighted PageRank, with respect to the threshold angle 45, based on the principle of every-best-fit and using the case of the Sydost region
(NOTE: for both PageRank and weighted PageRank, they have a series of PageRank scores with respect to different damping factor d values)

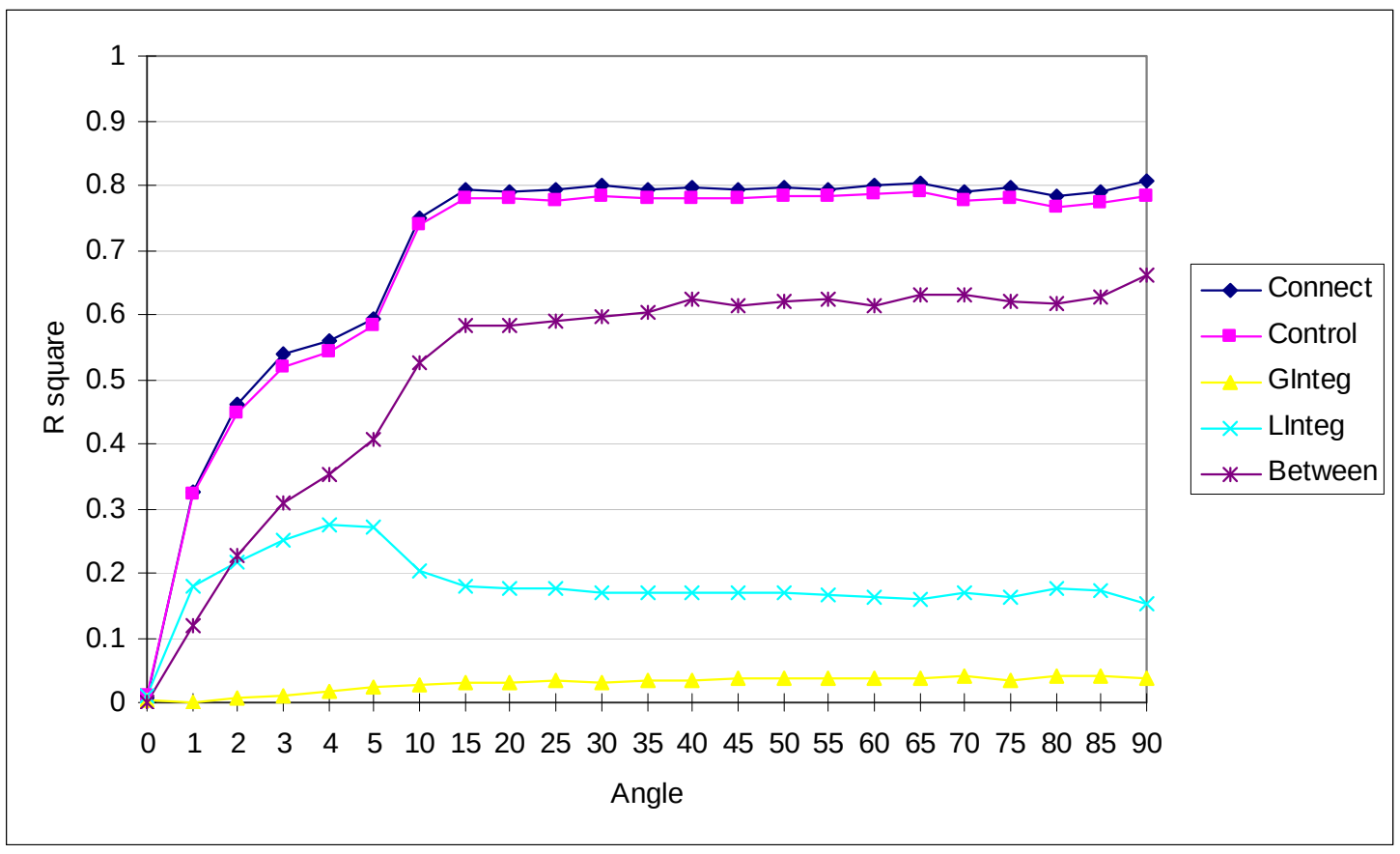

(a)

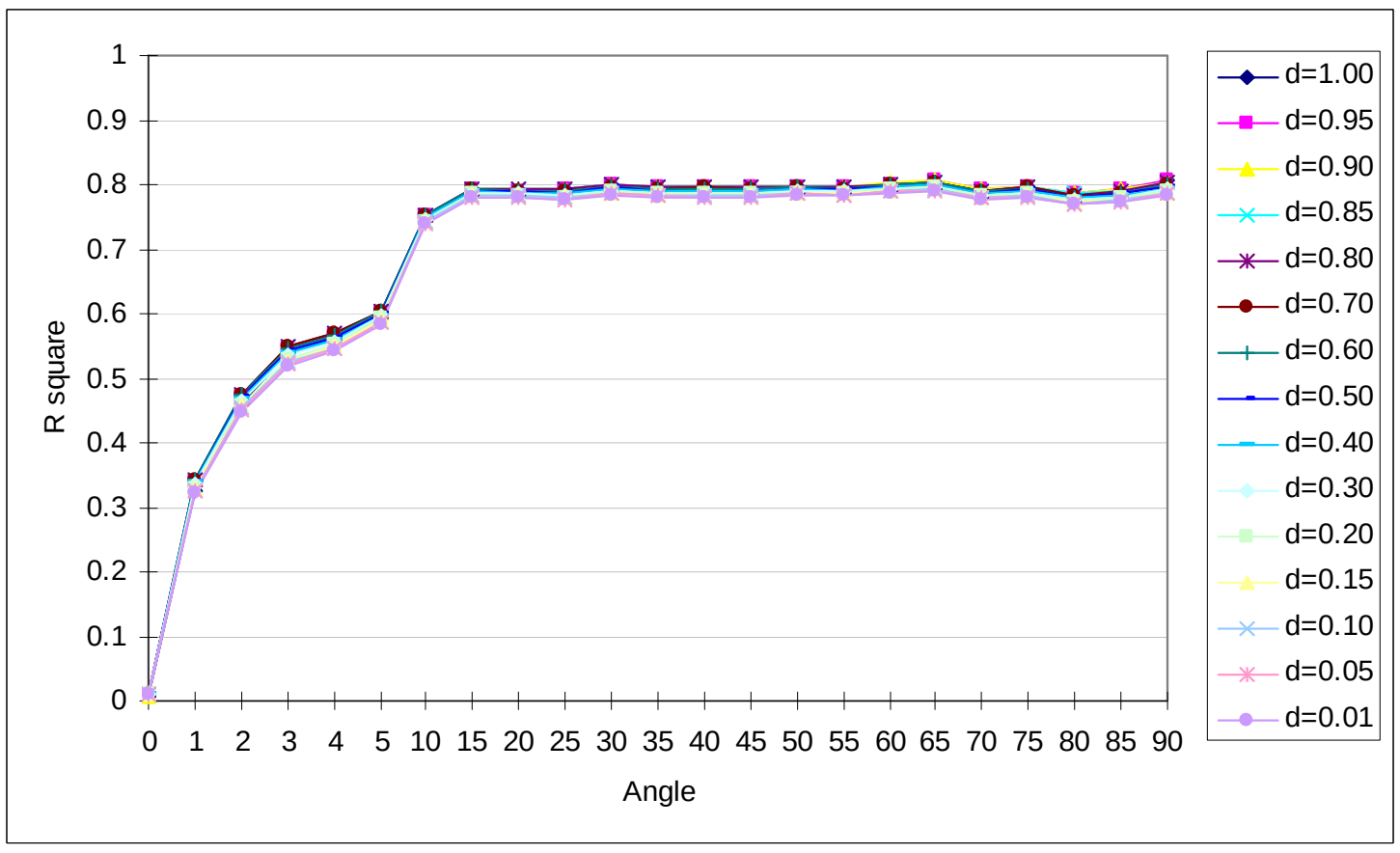

(b)

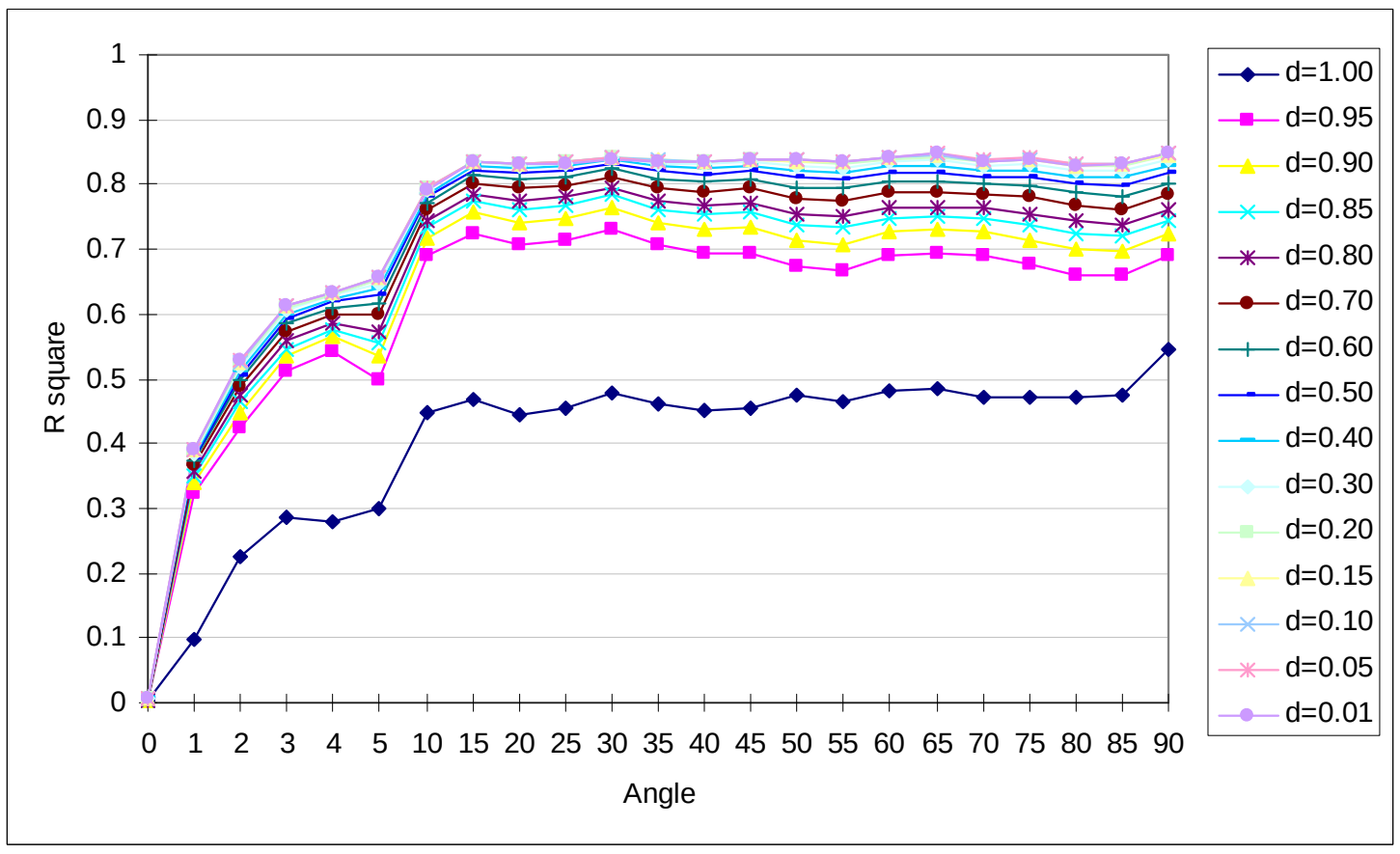

(c)

Figure 8: (color online) Correlation coefficient (R square) between traffic flow and (a) five centrality-based metrics, (b) PageRank and (c) weighted PageRank, with respect to the threshold angle, based on the principle of self-best-fit and using the case of the Sydost region
(NOTE: for both PageRank and weighted PageRank, they have a series of PageRank scores with respect to different damping factor d values. The curves are the averaged result of 20 experiments.)

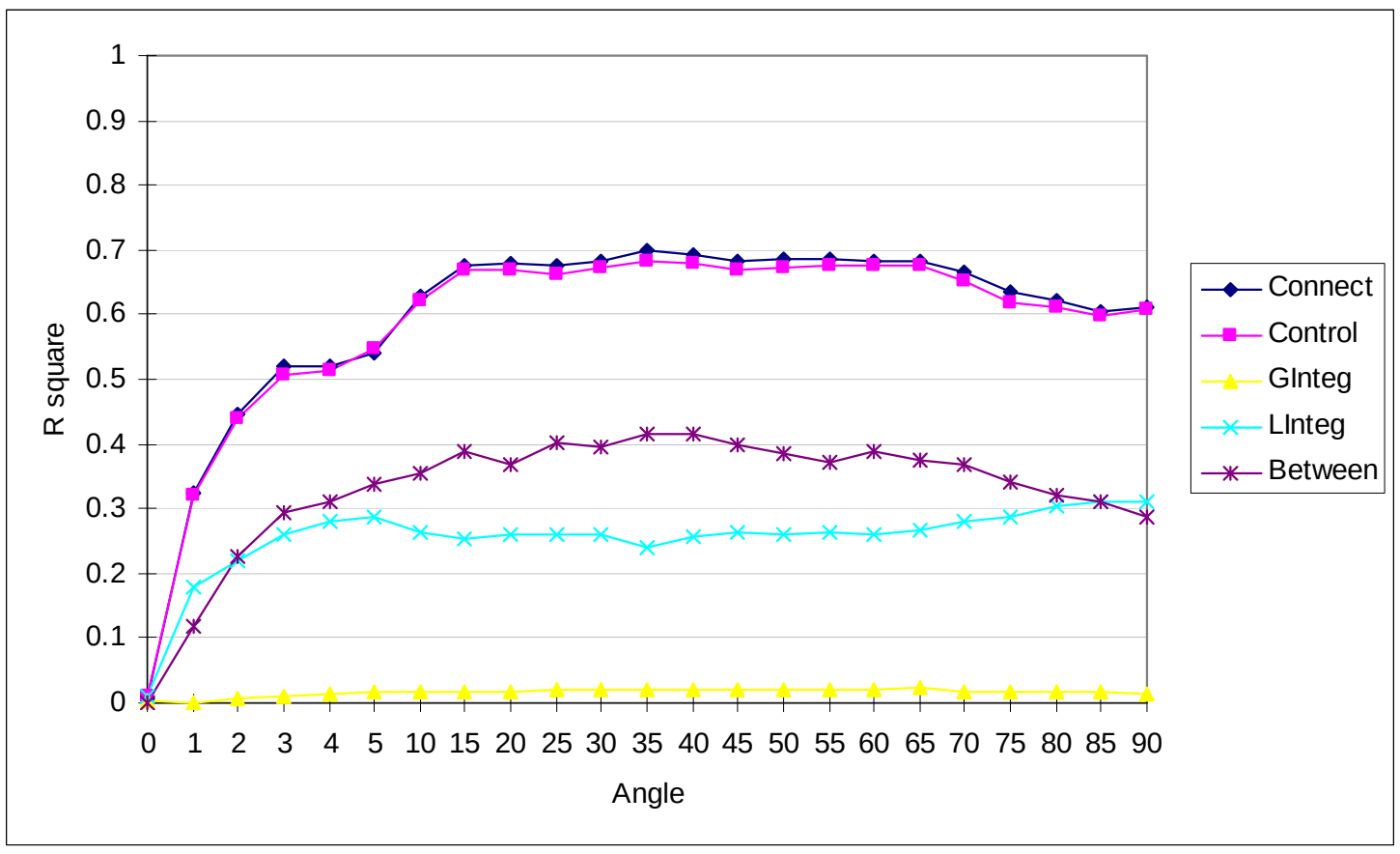

(a)

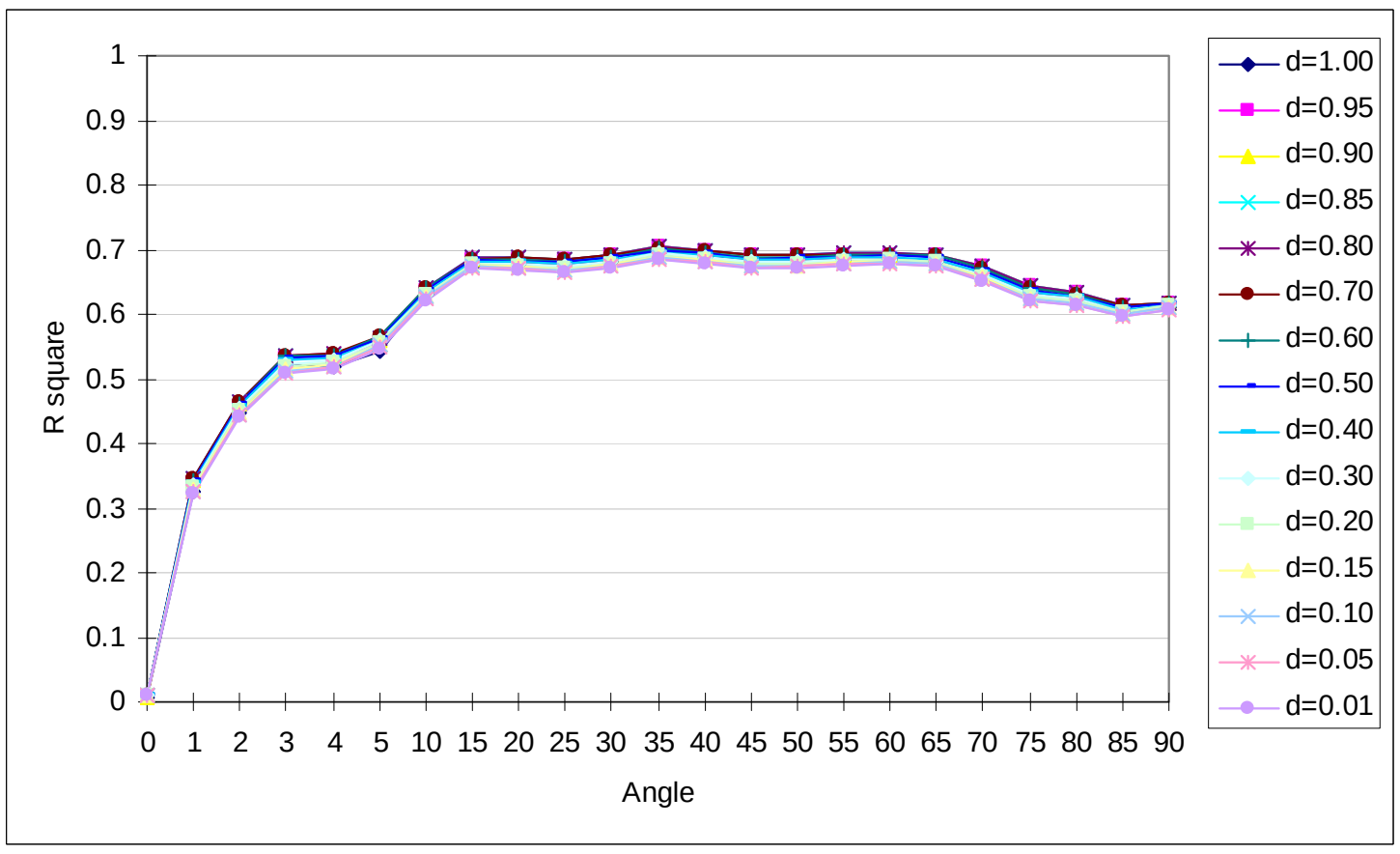

(b)

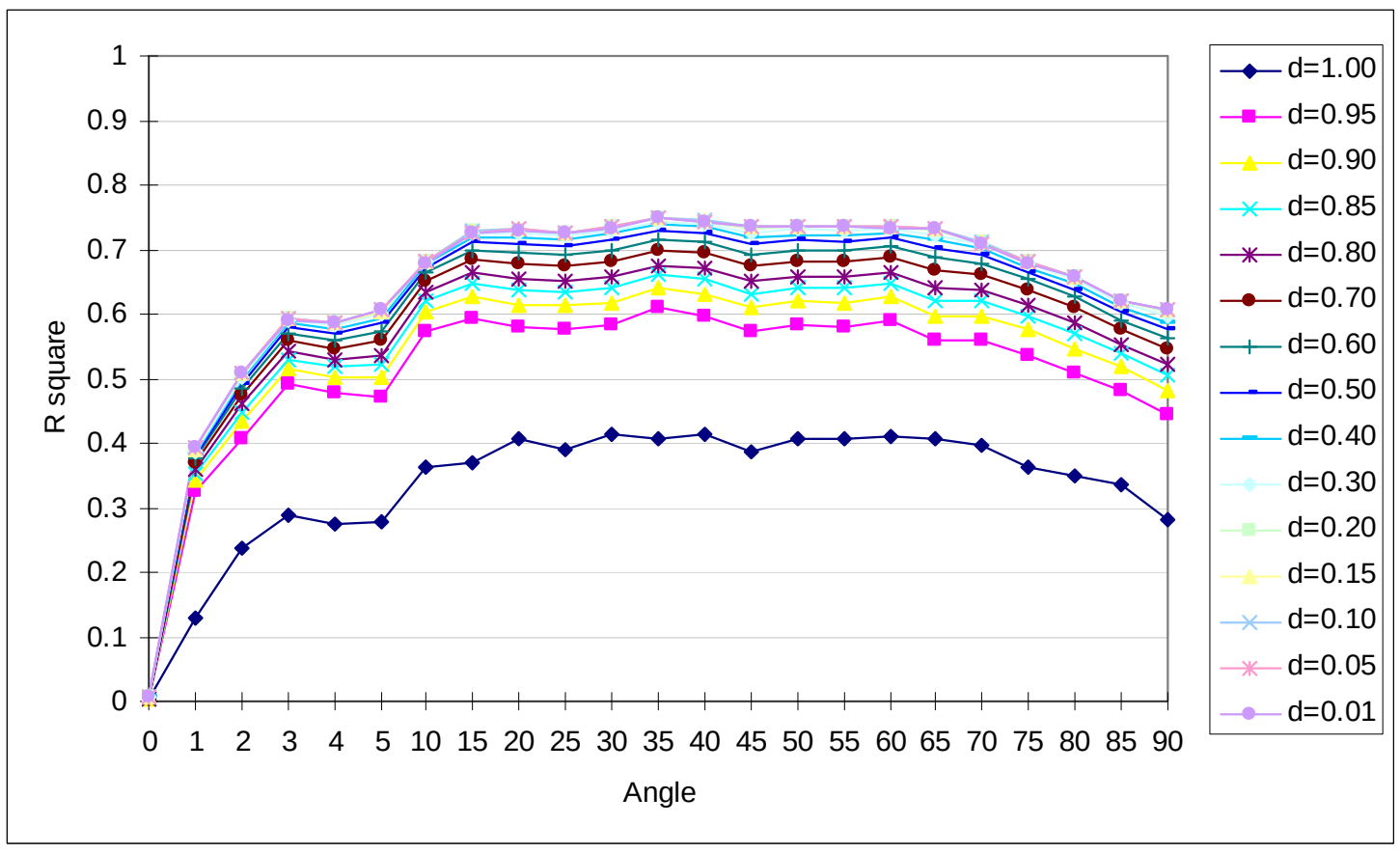

(c)

Figure 9: (color online) Correlation coefficient (R square) between traffic flow and (a) five centrality-based metrics, (b) PageRank and (c) weighted PageRank, with respect to the threshold angle 45, based on the principle of self-fit and using the case of the Sydost region
(NOTE: for both PageRank and weighted PageRank, they have a series of PageRank scores with respect to different damping factor d values. The curves are the averaged result of 20 experiments.)

Similar findings can be observed with the Gävle urban street network (Figures 10, 11 and 12). For instance, no metric-flow correlation exists at all for segments, but a significant correlation for streets. However, the correlation tends to become stable for PageRank metrics between degree 30 and 75, rather than between 15 and 90 in the previous case of nationwide networks. Weighted PageRank (when d = 0.95) is still the best metric (R square over 0.7) in terms of metric-flow correlation. It is followed by betweenness, whose R square is over 0.6. This result conforms to previous studies (Hillier and Iida 2005; Turner 2007). Again both local and global integrations are the poorest in terms of metric-flow correlation. More importantly, the principle self-best-fit is proved to be the best option. As for a possible reason, we will speculate on it later on.

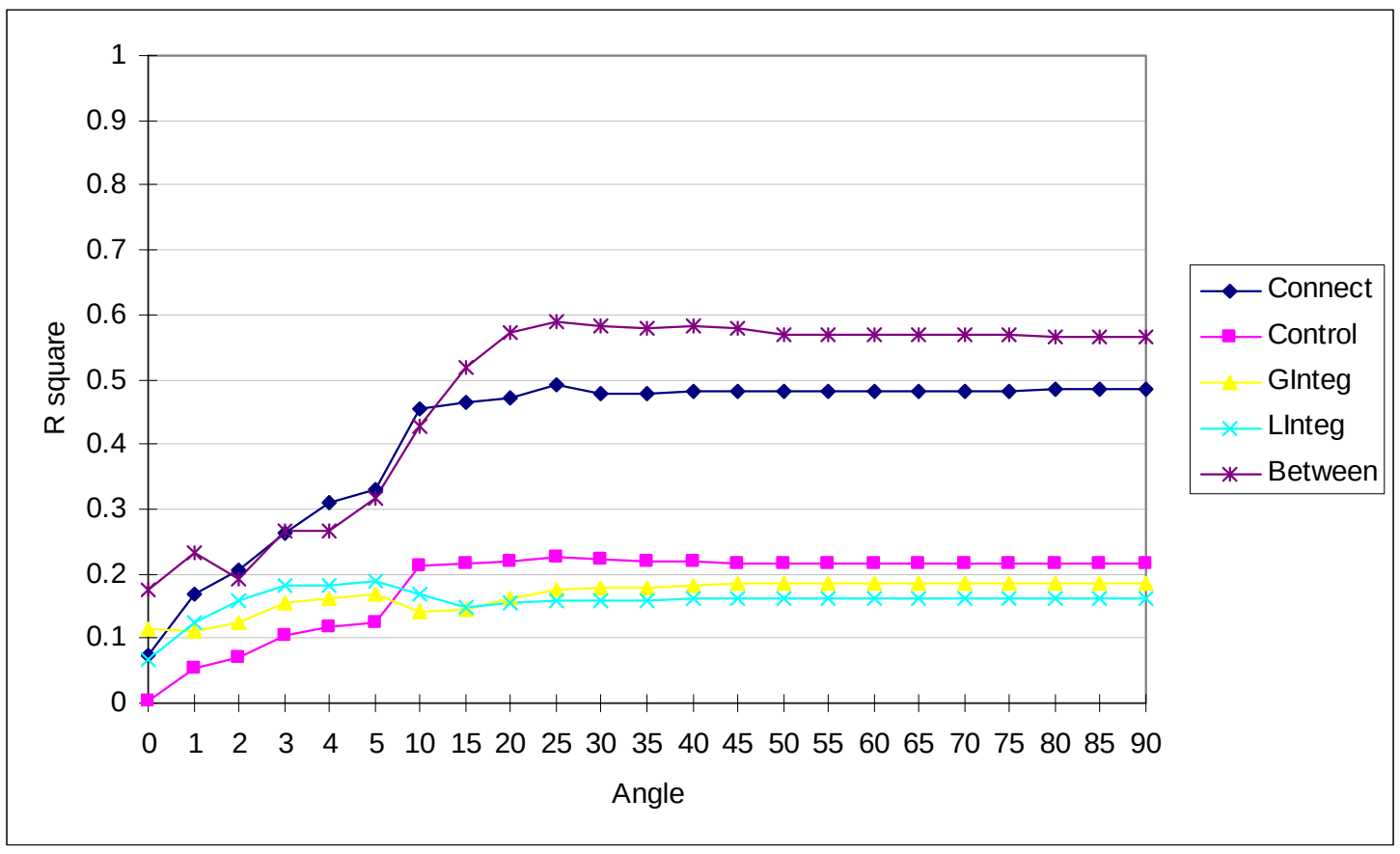

(a)

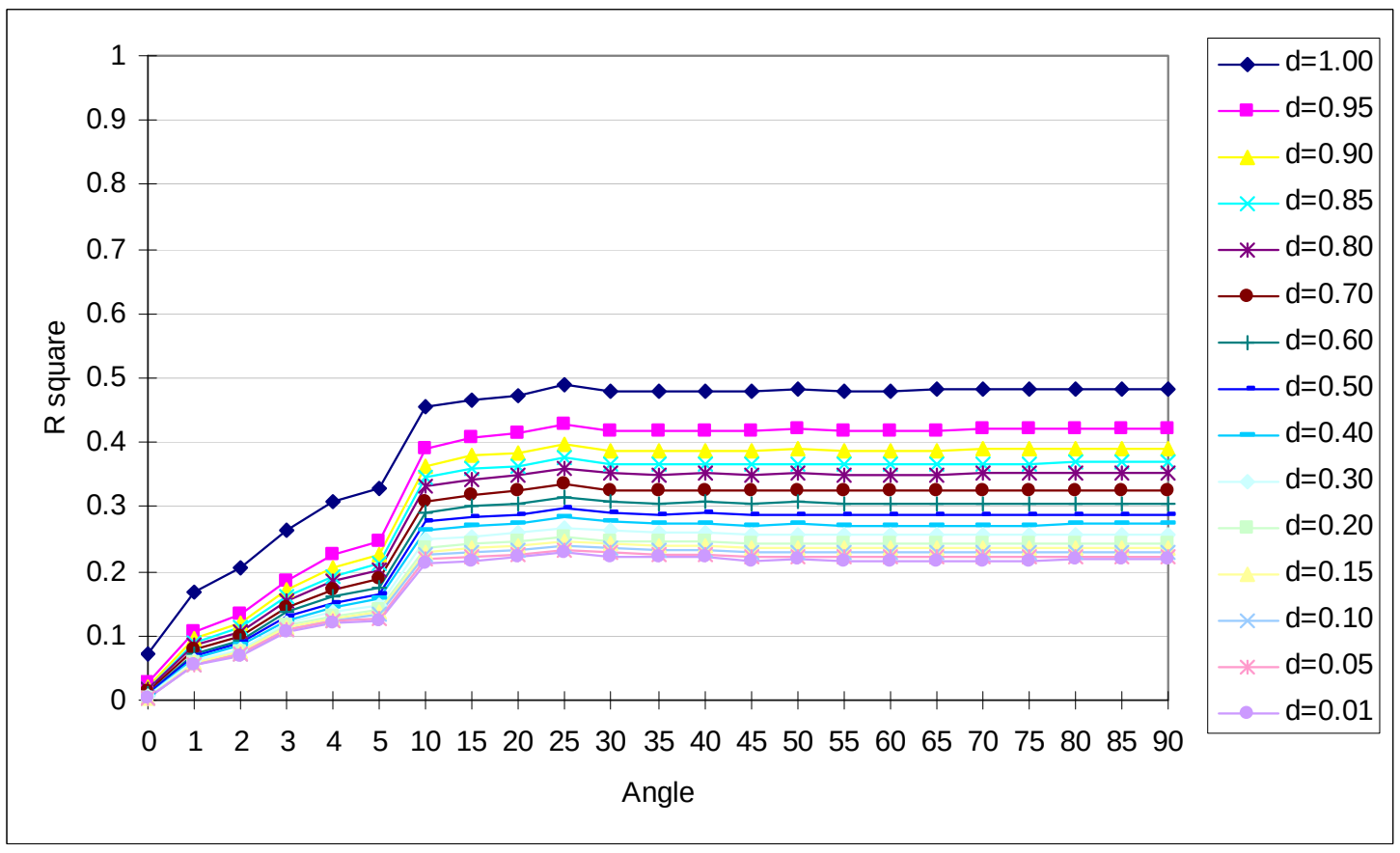

(b)

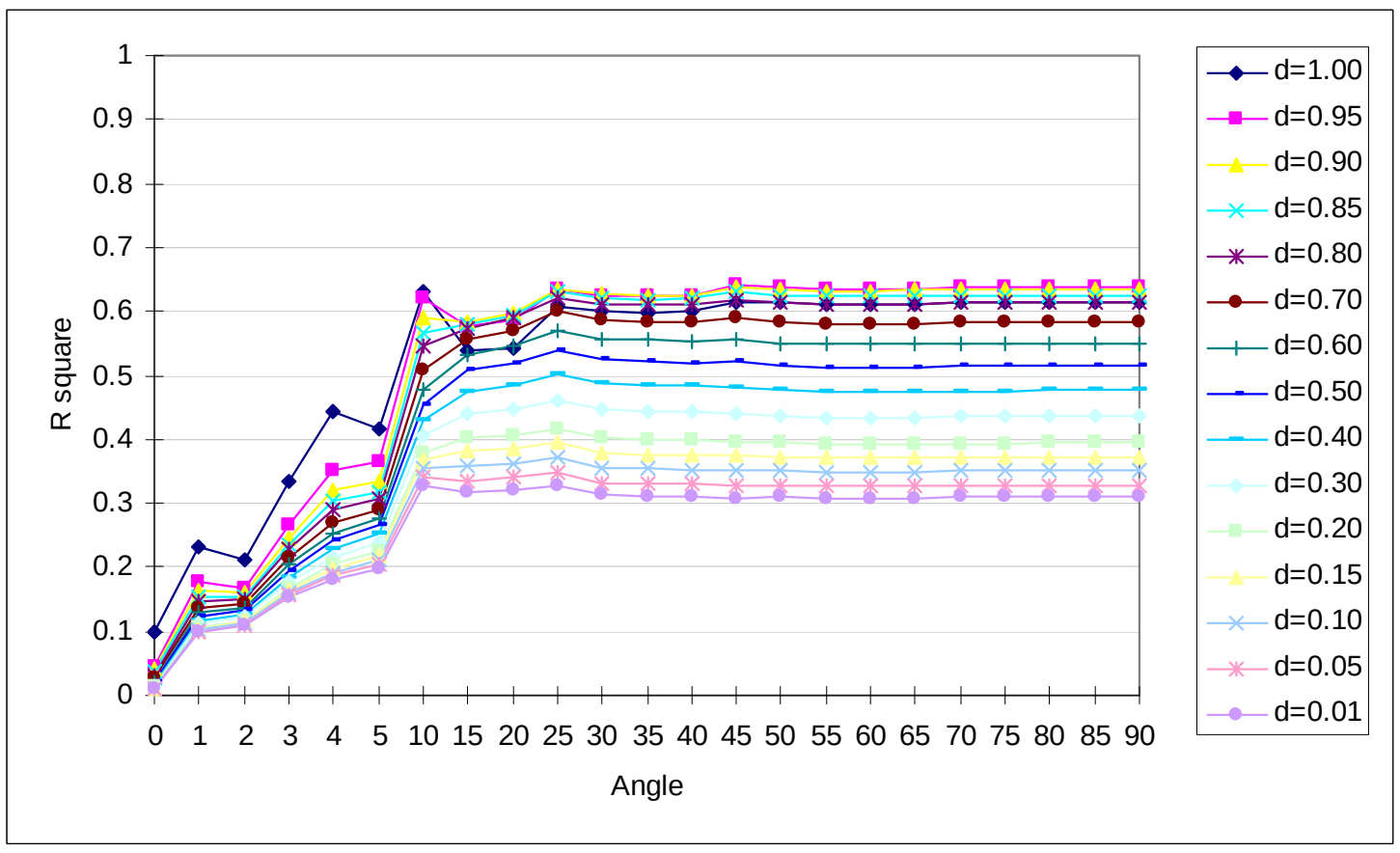

(c)

Figure 10: (color online) Correlation coefficient (R square) between traffic flow and (a) five centrality-based metrics, (b) PageRank and (c) weighted PageRank, with respect to the threshold angle 45, based on the principle of every-best-fit and using the case of the Gävle street network and one day traffic flow
(NOTE: for both PageRank and weighted PageRank, they have a series of PageRank scores with respect to different damping factor d values)

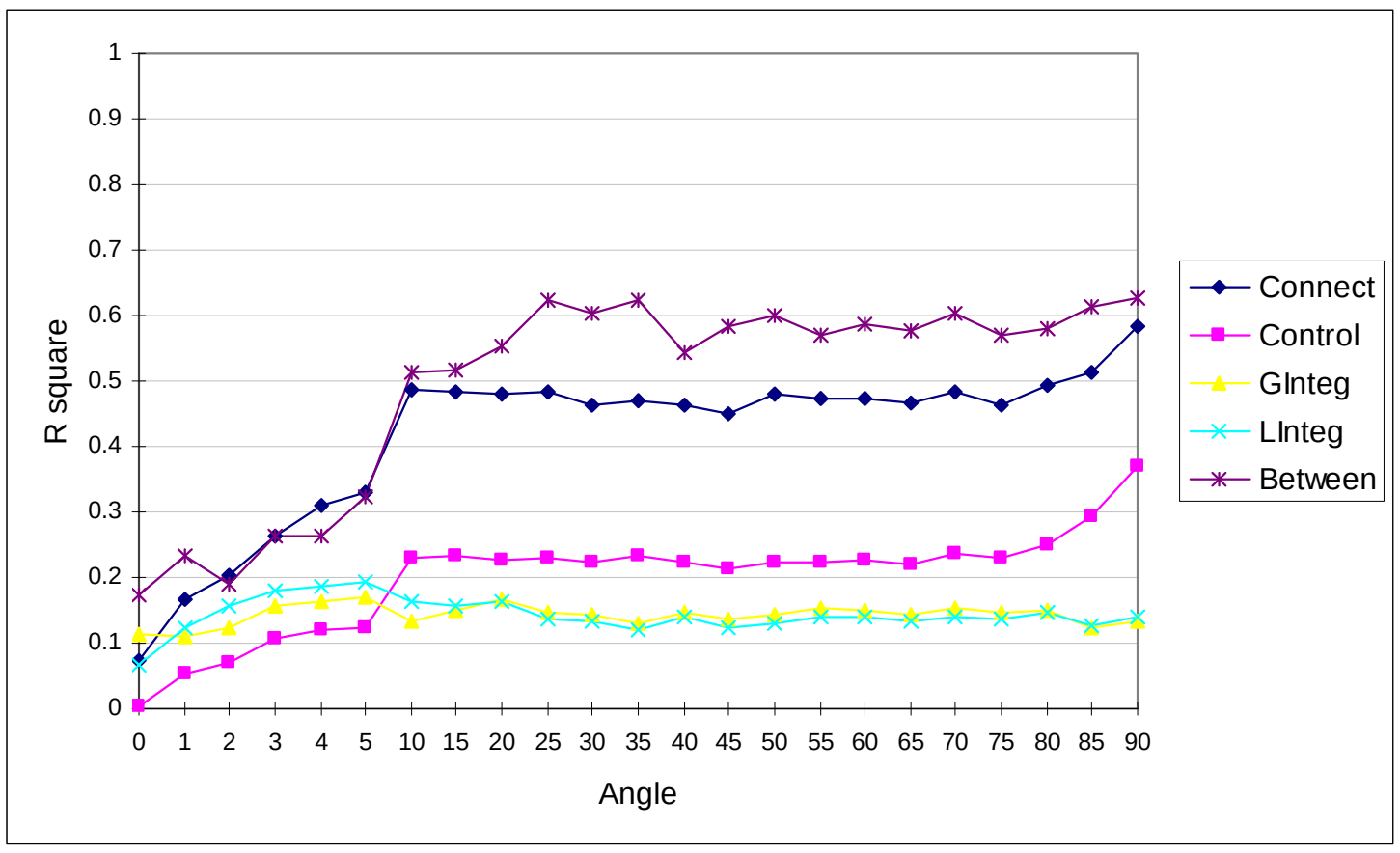

(a)

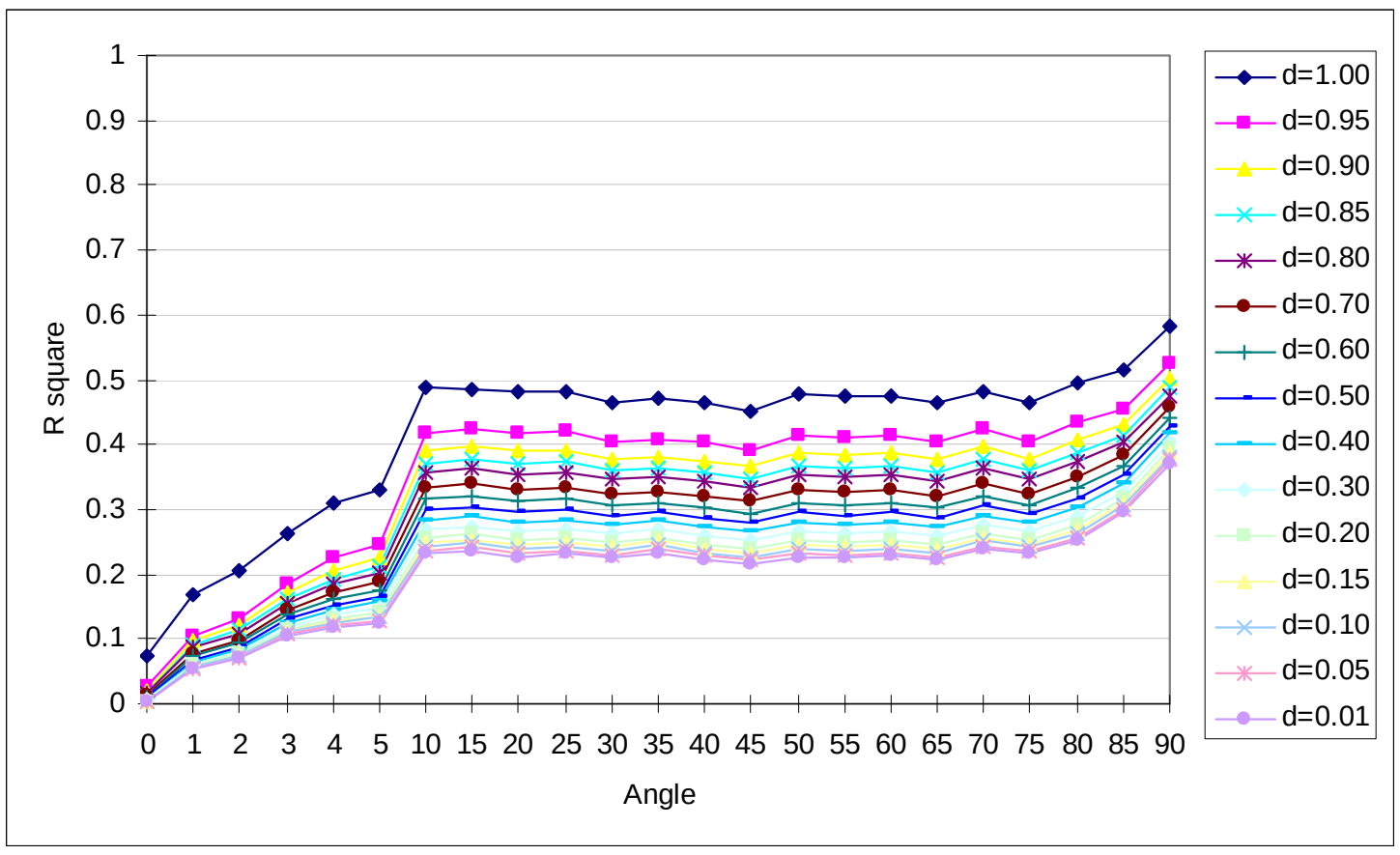

(b)

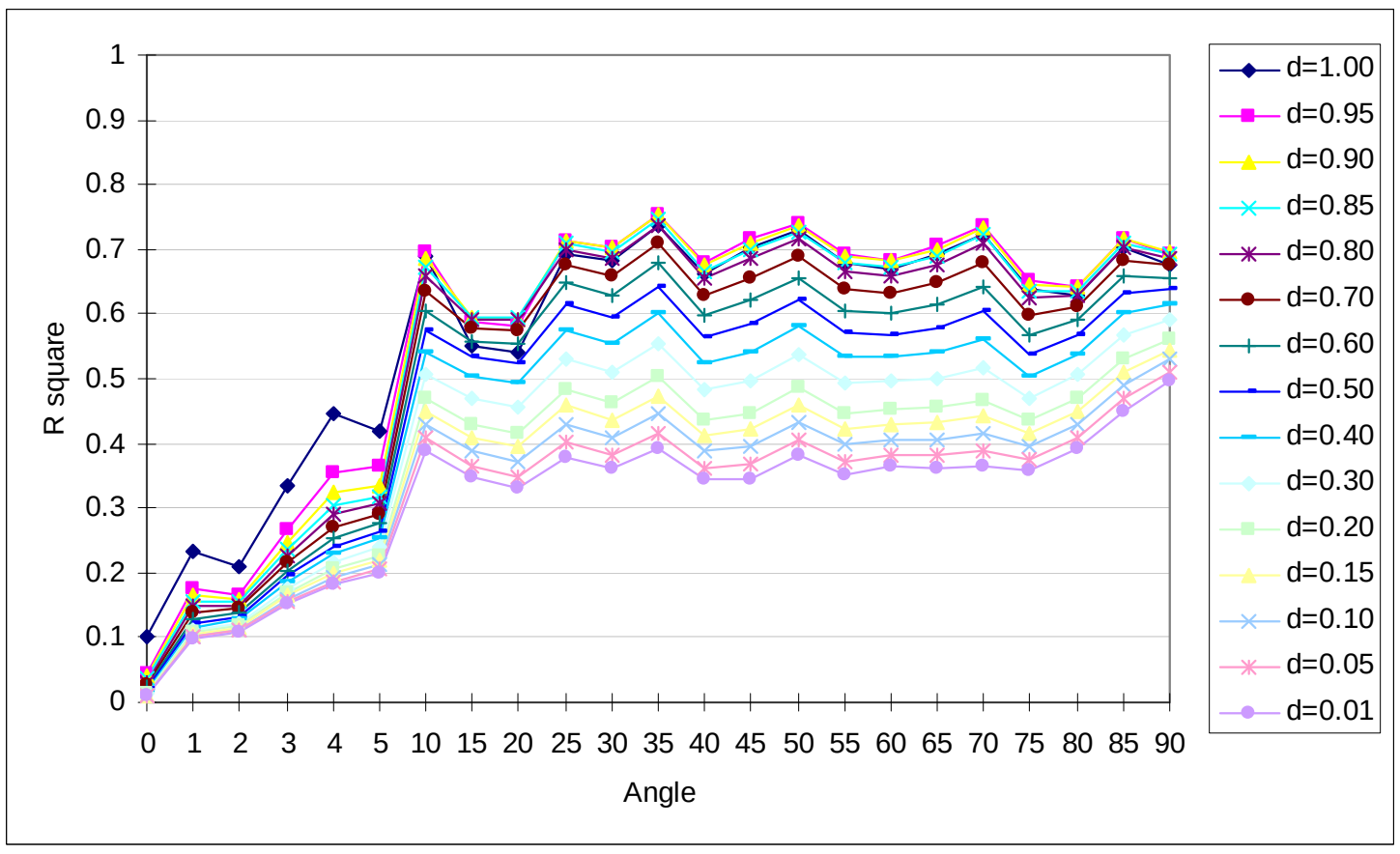

(c)

Figure 11: (color online) Correlation coefficient (R square) between traffic flow and (a) five centrality-based metrics, (b) PageRank and (c) weighted PageRank, with respect to the threshold angle 45, based on the principle of self-best- fit and using the case of the Gävle urban street network and one day traffic
(NOTE: for both PageRank and weighted PageRank, they have a series of PageRank scores with respect to different damping factor d values. The curves are the averaged result of 20 experiments.)

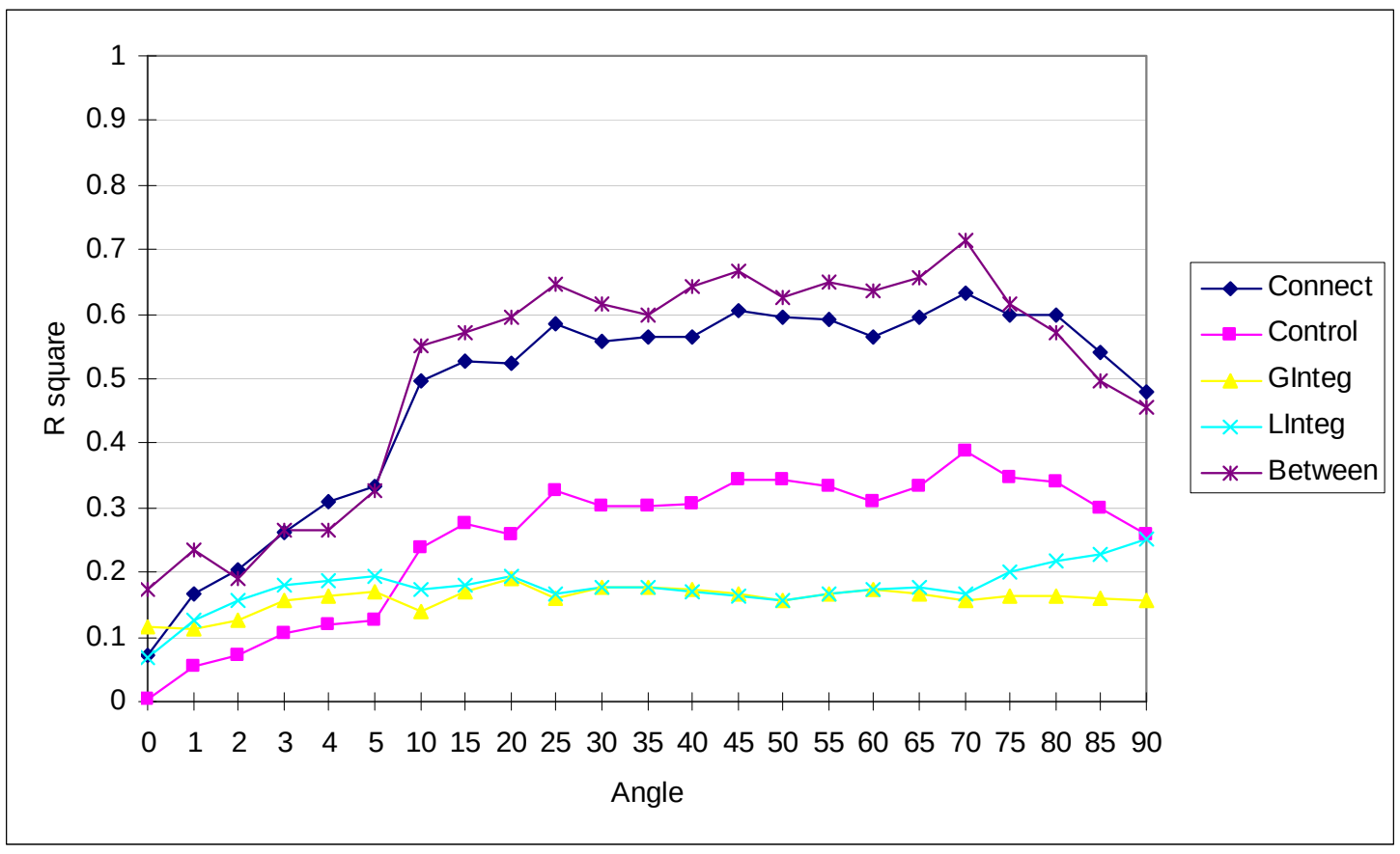

(a)

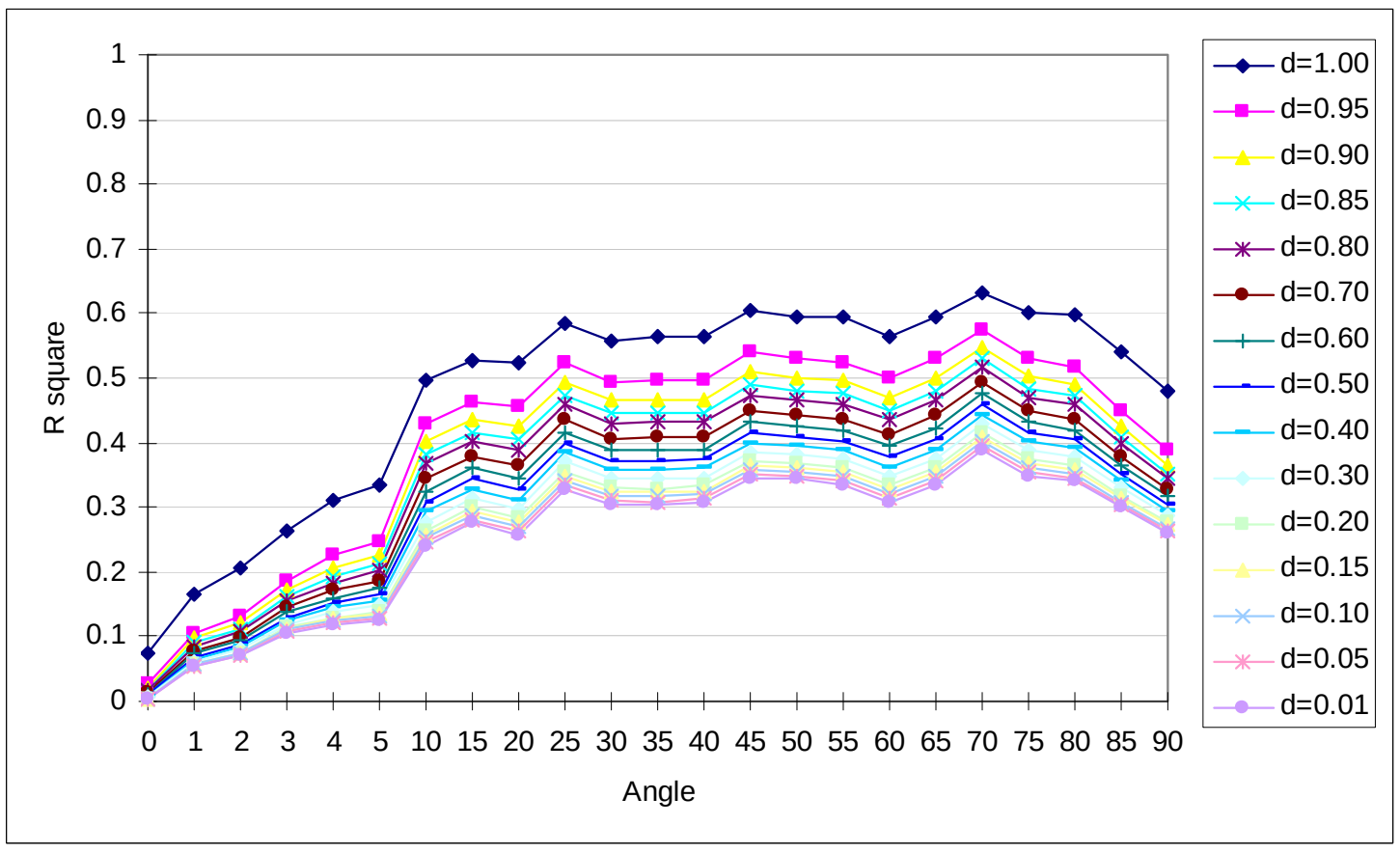

(b)

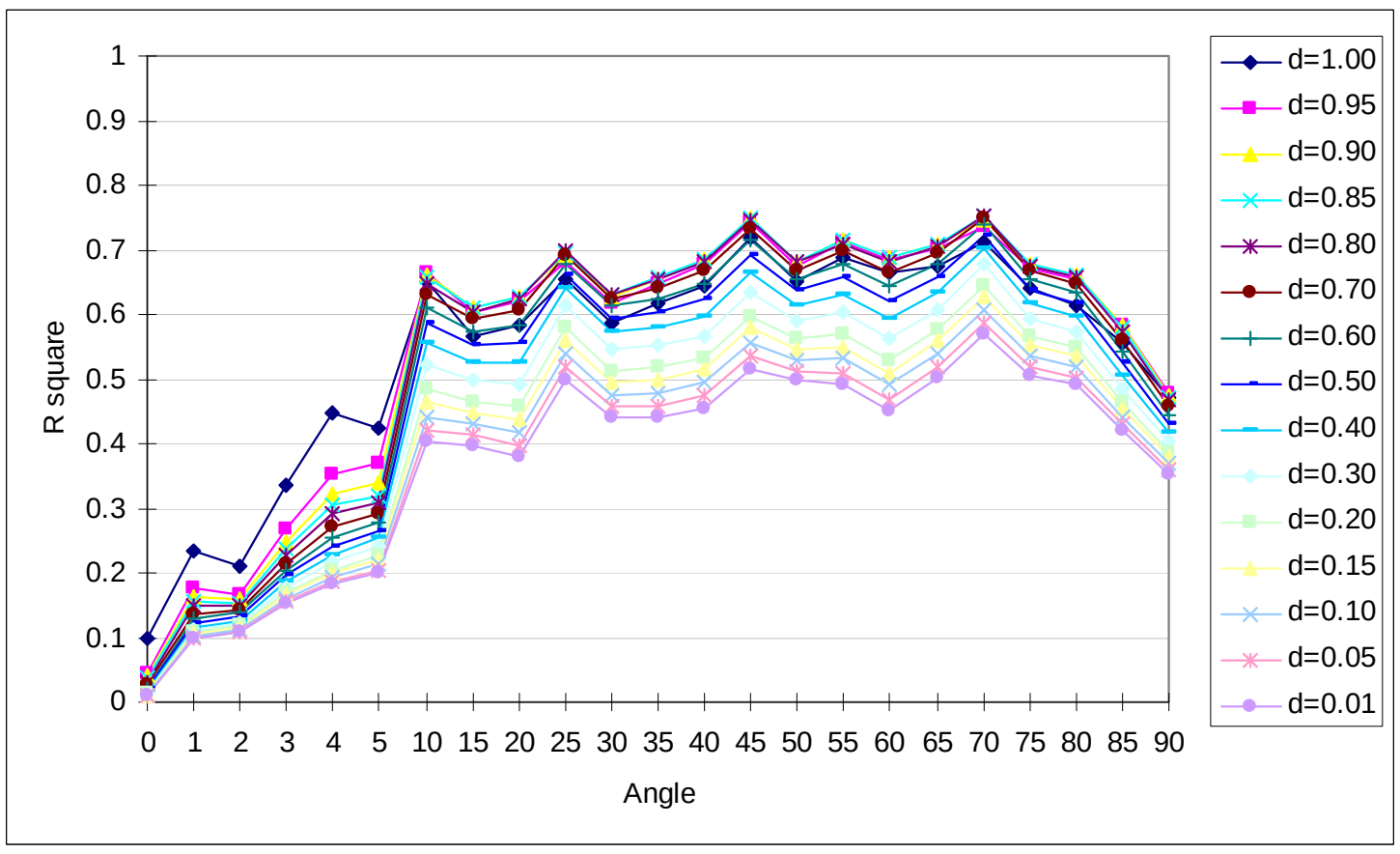

(c)

Figure 12: (color online) Correlation coefficient (R square) between traffic flow and (a) five centrality-based metrics, (b) PageRank and (c) weighted PageRank, with respect to the threshold angle, based on the principle of self-fit and using the case of the Gävle urban street network and one day traffic
(NOTE: for both PageRank and weighted PageRank, they have a series of PageRank scores with respect to different damping factor d values. The curves are the averaged result of 20 experiments.)

**4.3 Findings based on the point-based approach**

In this experiment, we adopt the point-based approach for forming connectivity graphs, and then assign point-based metrics by summation into individual roads. Surprisingly, both local and global integrations, previously demonstrated no scaling property using the line-based approach (Figures 5 and 6), exhibit a striking power law distribution (Figures 13 and 14).

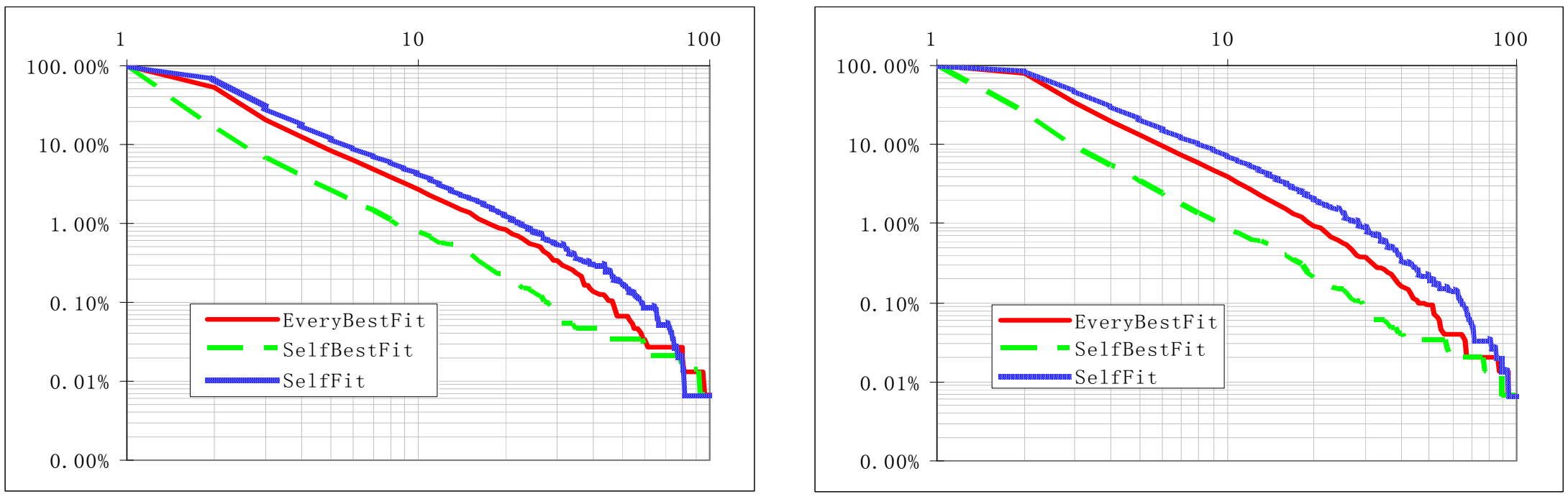

(a) (b)

Figure 13: Log-log plots of local (a) and global (b) integration using the point-based approach (the case of the entire nationwide road network)

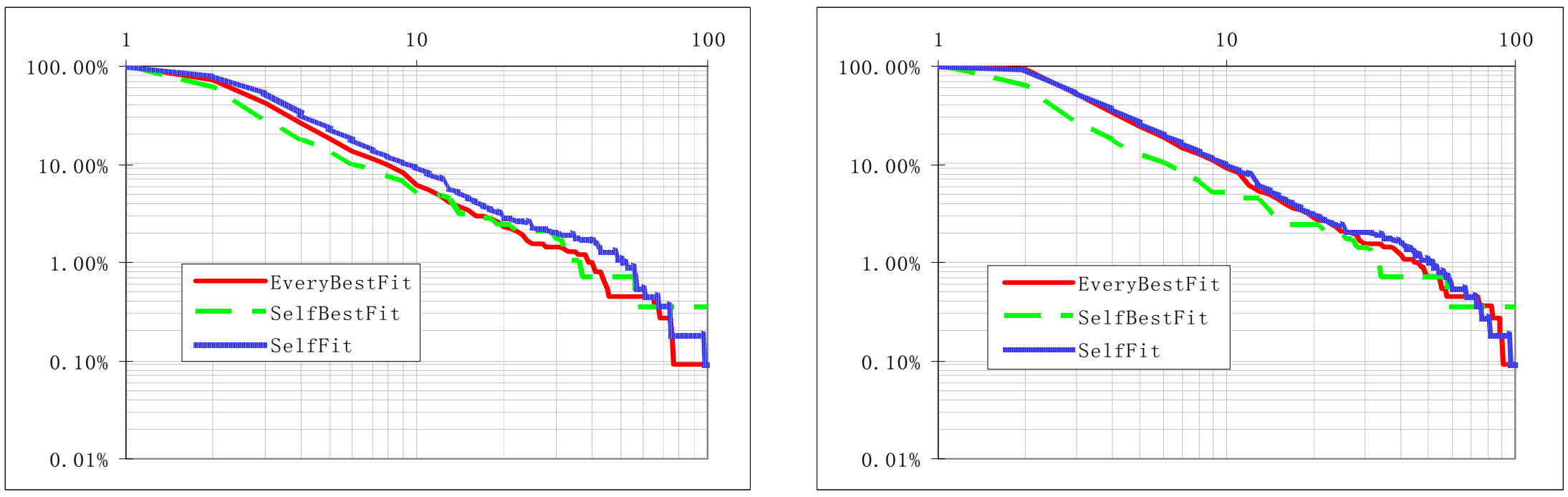

(a) (b)

Figure 14: (color online) Log-log plots of local (a) and global (b) integration using the point-based approach (the case of the Gävle street network)

There is a significant improvement for local and global integrations in terms of metric-flow correlation. There is no correlation (R square values less than 0.20) for local and global integrations with the line-based approach. However, R square values reach to around 0.8 for the case of Sydost (Figure 15a) when the point-based approach is adopted. A similar observation can be made for the case of Gävle (Figure 15b). To this point, we can remark that there is a significant relationship between scaling and metric-flow correlation. For instance, with the line-based approach both local and global integrations do not follow the scaling law (Figures 5 and 6), and there is nearly no metric-flow correlation for the integrations (Figures 7-12). However, in the space syntax community, it is commonly accepted that local and global integrations are the default indicators for traffic flow. It is absolutely not the case in our experiments based on the line-based approach. However, when we shift from the line-based to the point-based approach, local and global integrations become the default indicators for traffic. It appears a close relationship between the metrics' scaling distribution (Figures 13 and 14), and the metric-flow correlation (Figures 15 and 16).

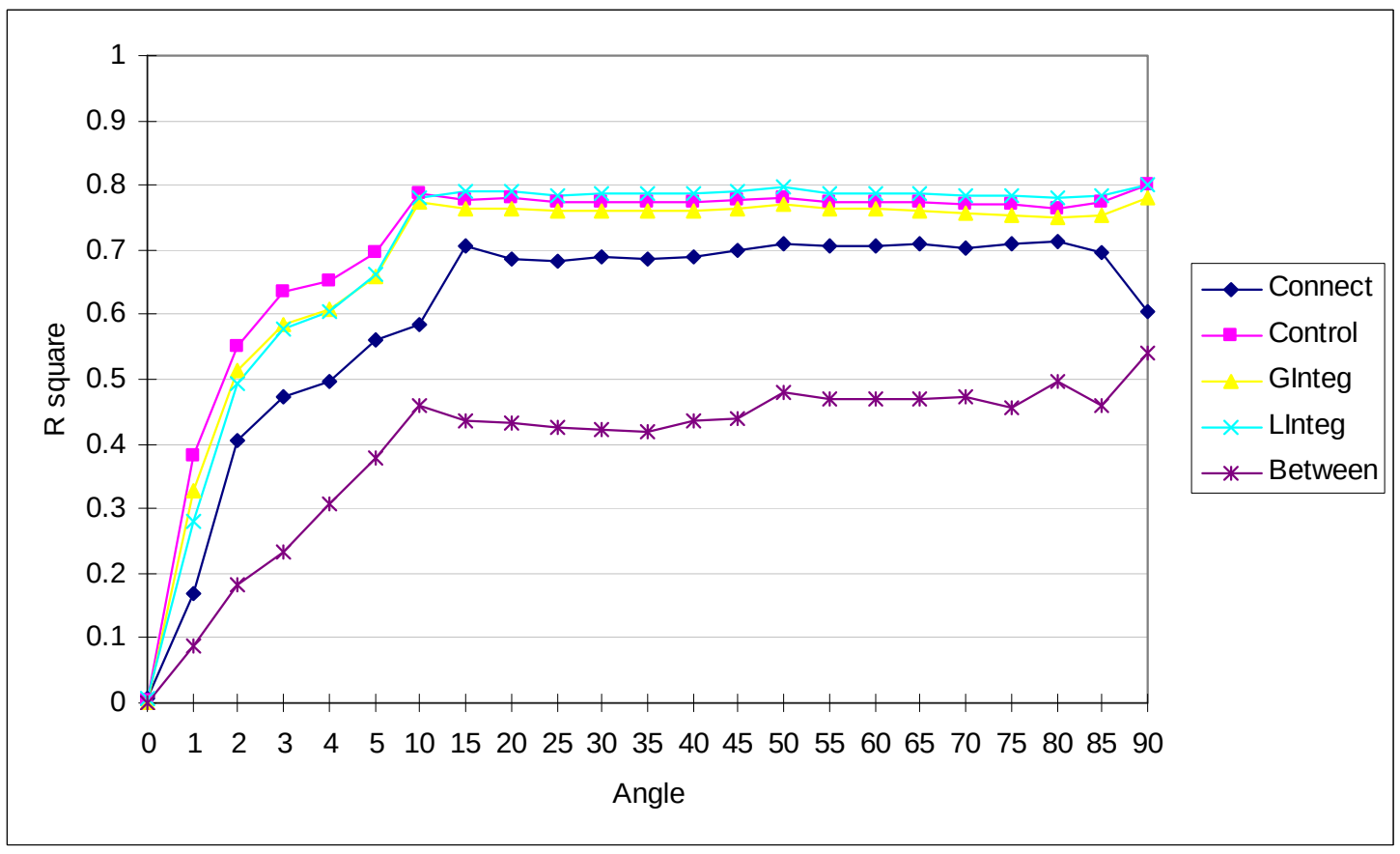

(a)

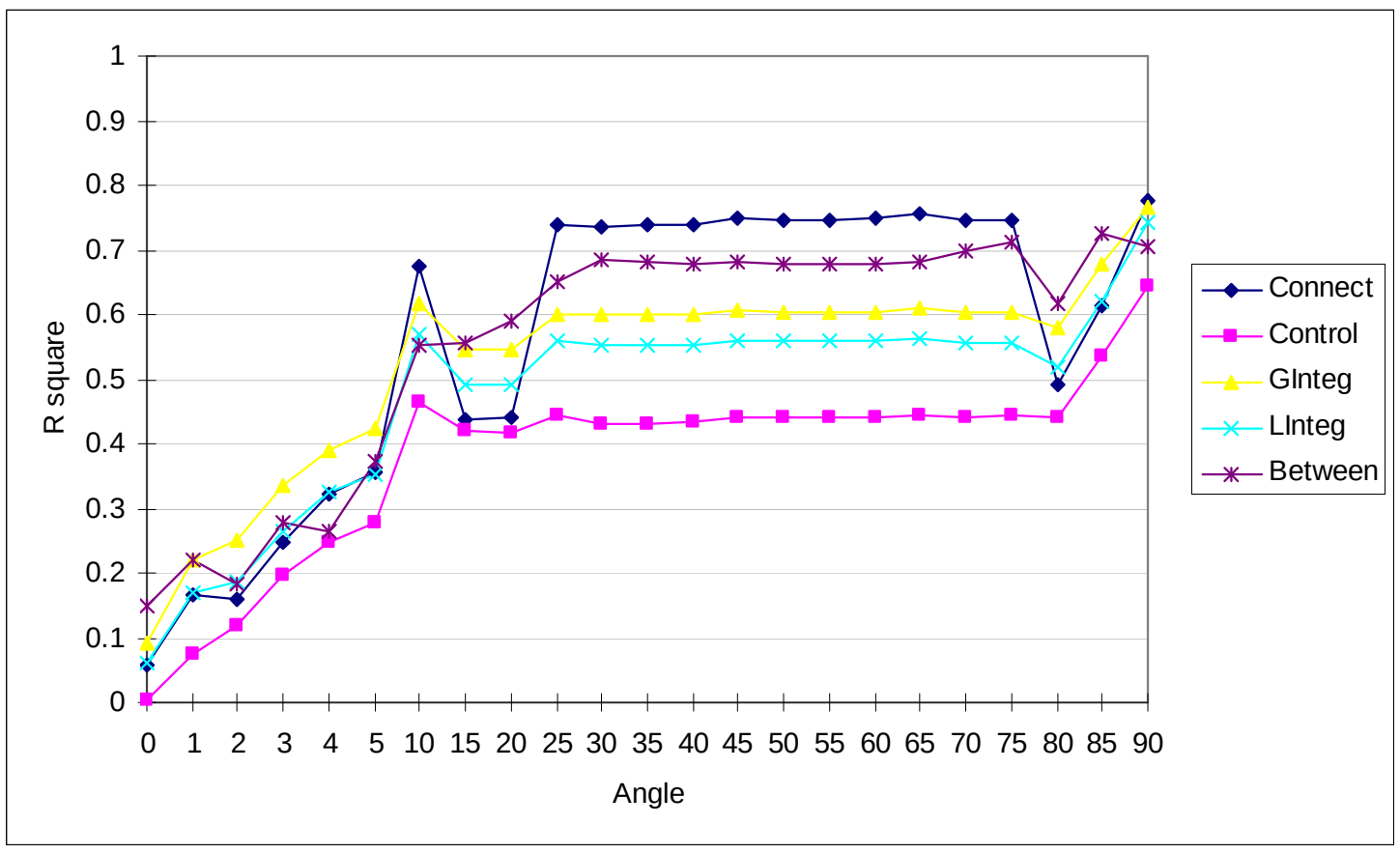

(b)

Figure 15: (color online) Correlation coefficient (R square) between traffic flow and point-based centrality metrics (a) the case of Syndost and (b) the case of Gävle (NOTE: local and global integrations in particular)

**5. Discussions on emergent properties of natural roads**

What we have found through the experiments can be considered to be emergent properties of natural roads, because they are not properties of the fundamental element, i.e., road segments. In the above experiments, we have illustrated that roads distinguish from segments in terms of the general behaviors. Roads are generated from segments by a self-organized process, but they demonstrate intelligence that underlying constituent segments lack. There is a striking zone between degree 30 and 75, where the emergent properties retain stable: (1) the number of natural roads remains stable, and (2) the metric-flow correlation does not alter much. The emergent properties remain valid for both nationwide and urban road networks. Some slight differences between nationwide and urban road networks do exist. Among others, the betweenness metric tends to be a good indicator for traffic flow in urban rather than nationwide settings. Under the line-based approach, traffic flow and all metrics except local and global integrations demonstrate the scaling property. However, while remaining unchanged for the scaling with all other metrics and traffic flow, local and global integrations exhibit the scaling (c.f. Figures 13 and 14, in comparison with Figures 5 and 6) when the point-based approach is adopted, more specifically, when point-based metrics are assigned by summation into individual roads. Although the underlying principle or mechanism still waits to be found, we try to justify our findings from the perspective of multi-agent systems (MAS) or complex adaptive systems (CAS) (Maes 1994; Johnson 2002).

The emergent properties found can be considered to be the outcome of interactions of individual segments from the bottom up. The segments, roads, and threshold angle (e.g., degree 45) can be compared to the sand grains, avalanches and slope in Bak's sandpile model (Bak, Tang and Wiesenfield 1987; Bak 1996). Segments can be regarded as multiple agents at the micro-scale, in which every individual segment interacts with its adjacent neighbors (at both ends) to form individual roads. The forming process can be regarded as a tracking process in which each segment at every junction point chooses one with the smallest angle to join. In the end, the formed roads meet the condition of energy minimization. It is a natural way of forming roads. In this regard, the formation of roads (Figure 16) resembles that of avalanches in Bak's sandpile model, where sand is added continuously, and a series of avalanches are generated as long as the slope of sand piles is greater than a threshold. Surprisingly, the size of roads, as well as that of avalanches, demonstrates a regularity of power law. Furthermore, the selfish oriented principles tend to capture traffic much better than the utopian one. This may sound counter intuitive, but it is exactly the diversity, one of the distinguished natures of CAS, that makes the difference. In other words, the more diverse the agents are, the more intelligent the CAS.

Roads can be regarded as multiple agents at the macro-scale, in which every road interacts with each other to form a connected whole. In the connected whole, all the roads collectively determine an individual's status. This is particularly true for PageRank metrics, which is justified by a federal system in which each agent (as a web page) casts a vote to determine an individual's status (Surowiecki 2004). More than that, an important page tends to have a higher weight than a less important page. Thus it is not a perfect democracy in the sense of one page one vote. In other words, not only popularity but also prestige determines the rankings in the whole. In addition, centrality metrics have also this nature of collective determination, although they are not as smart as PageRank metrics. Overall, the collectively determined metrics capture very well the traffic flow. This is another emergent property developed from the interactions of individual roads at the macro-scale.

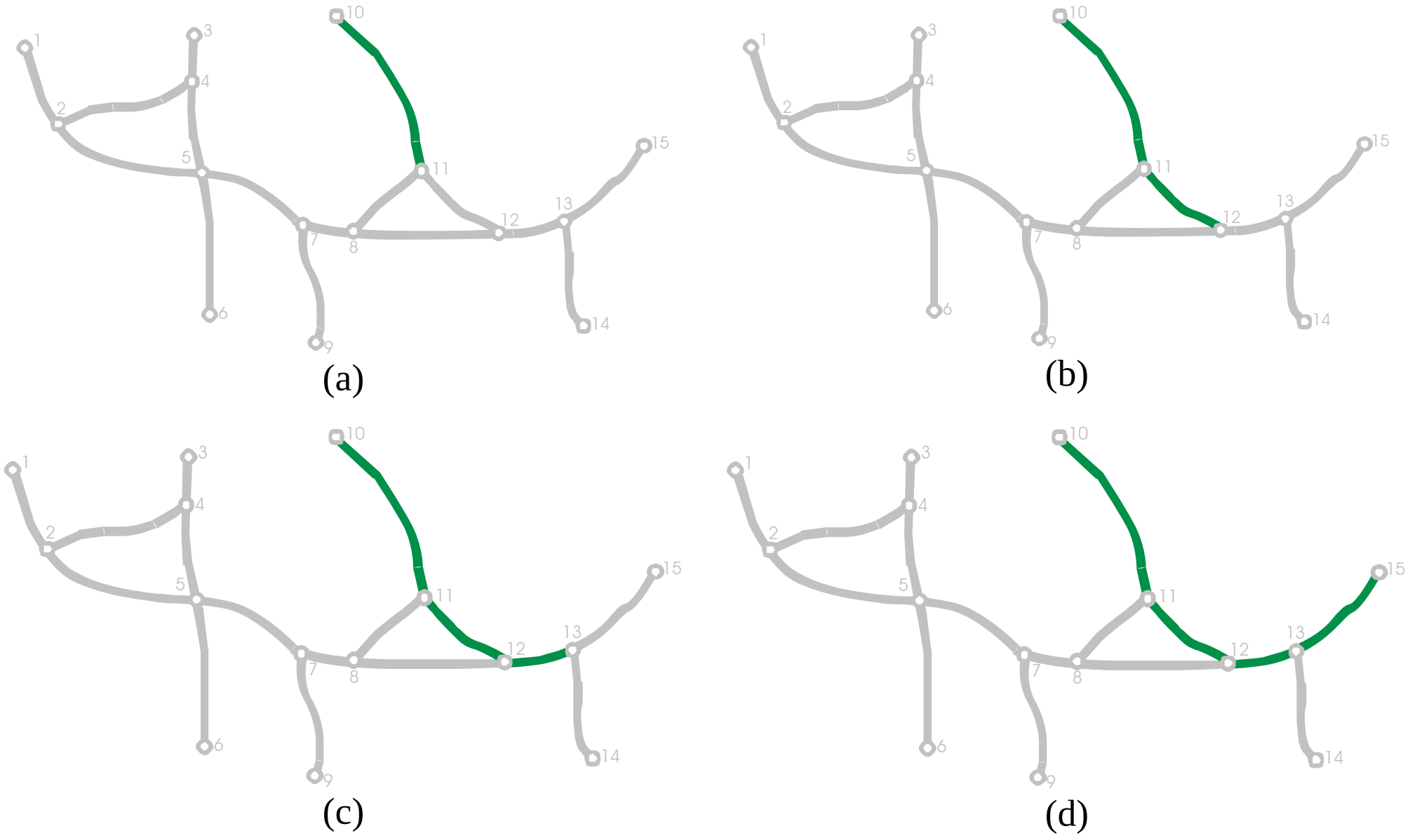

Figure 16: (color online) Formation of a natural road (green) in the sequence of (a), (b), (c) and (d) using the principle of self-best-fit

A third emergent property is with respect to the point-based approach. If we assign point-based metrics by summation into individual segments, then there would be no metric-flow correlation at all. However, things would be rather different if the point-based metrics are assigned into individual roads. First of all, local and global integrations exhibit a scaling property, whereas it lacks such a property in the line-based approach. Second, local and global integrations become the best metrics to capture traffic, while other metrics remain with no significant changes. To this point, we are still unable to provide a satisfactory justification as to why it is so. However, we conjecture that it is due to the Modifiable Areal Unit Problem (MAUP) (Openshaw 1984), which should be more properly named as the Modified Linear Unit Problem. MAUP refers to the fact that the aggregation units will affect statistics of spatial data, e.g., correlation relationships are strengthened by aggregations. As we have noticed, the point-based metrics assigned to segments have nearly no correlation, but tend to be highly correlated when the point-based metrics are assigned to roads. Clearly, metric-flow correlations are strengthened by aggregations from segments to roads. Whether this conjecture remains valid requires further investigation.

The emergent properties and intelligence demonstrated by natural roads provide telling evidence that cities are self-organized phenomena, which have life structure, as articulated by Alexander (2004) and Salingaros (2005). Linked to the empirical findings are also some fundamental, maybe philosophical, issues such as contradictions of uniformity and stupidity (of segments) versus diversity and intelligence (of roads), and unpredictability (of road length) versus predictability (of traffic). For instance, the principle self-best-fit based on natural selections makes natural roads more diverse than the other two principles. This is probably the reason why it is the best principle. The behavior change from segments to roads, in particular related to metric-flow correlation, sounds like a phase transition. This transition can be compared to that from ants to colonies, and sand grains to avalanches. These issues are fundamental to many complexity systems in nature or society, which deserve further research.

## 6. Conclusion

We studied road networks from the perspective of complex networks by concentrating on the sensitivity issues with respect to join principles, the damping factors with PageRank metrics, and the difference between line-based and point-based approaches. Using massive road networks and traffic flow data, we found that (1) there exists a tipping point from segment-based to road-based network topology in terms of correlation between ranking metrics and traffic flow, (2) the correlation is significantly improved if a selfish rather than utopian strategy is adopted in forming the self-organized natural roads, and (3) point-based metrics assigned by summation into individual roads tend to have a much better correlation with traffic flow than line-based metrics, and this is particularly true for both local and global integrations. In addition, we found that weighted PageRank with an appropriate *d* factor setting tends to be one of the best metrics for correlating or predicting traffic flow. In comparison with line-based and point-based approaches, the point-based one tends to be the best option.

We have tried to put natural roads in analogue with many complex phenomena such as ants/colonies and sand grains/avalanches, which demonstrate emergent properties and a universal regularity of power law distribution. We illustrated various emergent properties developed from roads and road network topology. Our study sheds substantial insights into the understanding of road networks. Road networks, although artifacts in nature, can be compared with biological entities, which exhibit complexity that is developed from the interaction of individuals, thus bottom-up in nature.

**Acknowledgements**
This research was financially supported by research grants from the Hong Kong Polytechnic University, and the Swedish Research Council FORMAS. The data on nationwide road networks and relevant traffic flow are provided by the Swedish Road Administration (Vägverket). In this regard, Michael Westin from Vägverket Region Mitt deserves our special thanks for his assistance in preparing and transferring the datasets. The Gävle street network is provided by Gävle city, and GPS data are provided kindly by the Gävle taxi company (TAXI Stor och Liten) in Sweden. In addition, the first author is grateful to Xintao Liu and Hong Zhang for their help in drawing Figure 1 and patience in discussing this paper while it was prepared for publication.

**References**
Alexander C. (2004), *The Nature of Order*, Center for Environmental Structure: Berkeley, CA.
Bak P. (1996), *How Nature Works: the science of self-organized criticality*, Springer-Verlag: New York.
Bak P., Tang C. and Wiesenfield K. (1987), Self-organized criticality: an explanation of 1/f noise, *Physical Review Letters*, 59, 381 – 384.
Batty M. (2004), A new theory of space syntax, *CASA working paper #75*.
Freeman L. C. (1979), Centrality in social networks: conceptual clarification, *Social Networks*, 1, 215 - 239.
Gastner M.T. and Newman M.E.J. (2006), The spatial structure of networks, *The European Physical Journal B*, 49, 247–252.
Hillier B. and Hanson J. (1984), *The Social Logic of Space*, Cambridge University Press: Cambridge.
Hillier B. and Iida S. (2005), Network and psychological effects in urban movement, In: Cohn A.G. and Mark D. M. (eds.), *Proceedings of the International Conference on Spatial Information Theory, COSIT 2005*, Ellicottsville, N.Y., U.S.A., September 14-18, 2005, Springer-Verlag: Berlin, 475-490.
Jiang B. (2006), Ranking spaces for predicting human movement in an urban environment, to appear in *International Journal of Geographical Information Science*, xx, xx – xx, *Preprint, arxiv.org/physics/0612011/*.
Jiang B. (2007), A topological pattern of urban street networks: universality and peculiarity, *Physica A*, 384, 647 – 655.
Jiang B. and Claramunt C. (2002), Integration of space syntax into GIS: new perspectives for urban morphology, *Transactions in GIS*, 6(3), 295-309.
Jiang B. and Claramunt C. (2004), Topological analysis of urban street networks, *Environment and Planning B: Planning and Design*, 31, 151- 162.
Jiang B. and Liu C. (2007), Street-based topological representations and analyses for predicting traffic flow in GIS, to appear in *International Journal of Geographic Information Science*, xx, xx – xx, *preprint, arxiv.org/abs/0709.1981*.
Johnson S. (2002), *Emergence: the connected lives of ants, brains, cities and software*, TOUCHSTONE: New York.
Langville A. N. and Meyer C. D. (2006), *Google's PageRank and Beyond: the science of search engine rankings*, Princeton University Press: Princeton, N.J.
Maes P. (1994), Modeling adaptive autonomous agents, *Artificial Life*, 1, 135 – 162.
Openshaw, S. (1984), *The Modifiable Areal Unit Problem*, Geo Book: Norwich.
Page L. and Brin S. (1998), The anatomy of a large-scale hypertextual Web search engine, *Proceedings of the Seventh International Conference on World Wide Web*, 7, 107-117.
Porta S., Crucitti P. and Latora V. (2006), The network analysis of urban streets: a dual approach, *Physica A*, 369, 853 – 866.
Salingaros N. A. (2005), *Principles of Urban Structure*, Techne: Delft.
Surowiecki J. (2004), *The Wisdom of Crowds: why the many are smarter than the few*, ABACUS: London.
Taxi (2007), Taxi Stor och Liten, www.taxi107000.com

Thomson R. C. (2003), Bending the axial line: smoothly continuous road centre-line segments as a basis for road network analysis, in Hanson, J. (ed.), *Proceedings of the Fourth Space Syntax International Symposium*, University College London, London.

Turner A. (2007), From axial to road-centre lines: a new representation for space syntax and a new model of route choice for transport network analysis, *Environment and Planning B: Planning and Design*, 34(3), 539–555.

Xing W. and Ghorbani A. (2004), Weighted PageRank algorithm, *Second Annual Conference on Communication Networks and Services Research CNSR'04*, Fredericton, N.B. Canada, 305 – 314.

**Appendix A: Matrices derived from the notational road network in Figure 1**

With reference to Figure 1 in the main text of this paper, we derived various matrices, representing different networks or graphs. First, Point-Point Distance Matrix (PPDM) is a matrix, a two-dimensional array, containing the distances of a set of points, which are road junctions or ends.

$$PPDM = \left[\begin{array}{cccccccccccccccc} & 1 & 2 & 3 & 4 & 5 & 6 & 7 & 8 & 9 & 10 & 11 & 12 & 13 & 14 & 15 \\ 1 & 0 & x & 0 & 0 & 0 & 0 & 0 & 0 & 0 & 0 & 0 & 0 & 0 & 0 & 0 \\ 2 & 0 & 0 & 0 & x & x & 0 & 0 & 0 & 0 & 0 & 0 & 0 & 0 & 0 & 0 \\ 3 & 0 & 0 & 0 & x & 0 & 0 & 0 & 0 & 0 & 0 & 0 & 0 & 0 & 0 & 0 \\ 4 & 0 & x & 0 & 0 & x & 0 & 0 & 0 & 0 & 0 & 0 & 0 & 0 & 0 & 0 \\ 5 & 0 & x & 0 & x & 0 & 0 & 0 & 0 & 0 & 0 & 0 & 0 & 0 & 0 & 0 \\ 6 & 0 & 0 & 0 & 0 & x & 0 & 0 & 0 & 0 & 0 & 0 & 0 & 0 & 0 & 0 \\ 7 & 0 & 0 & 0 & 0 & x & 0 & 0 & x & x & 0 & 0 & 0 & 0 & 0 & 0 \\ 8 & 0 & 0 & 0 & 0 & 0 & 0 & x & 0 & 0 & 0 & x & x & 0 & 0 & 0 \\ 9 & 0 & 0 & 0 & 0 & 0 & 0 & x & 0 & 0 & 0 & 0 & 0 & 0 & 0 & 0 \\ 10 & 0 & 0 & 0 & 0 & 0 & 0 & 0 & 0 & 0 & 0 & x & 0 & 0 & 0 & 0 \\ 11 & 0 & 0 & 0 & 0 & 0 & 0 & 0 & x & 0 & x & 0 & x & 0 & 0 & 0 \\ 12 & 0 & 0 & 0 & 0 & 0 & 0 & 0 & x & 0 & 0 & x & 0 & x & 0 & 0 \\ 13 & 0 & 0 & 0 & 0 & 0 & 0 & 0 & 0 & 0 & 0 & 0 & x & 0 & x & x \\ 14 & 0 & 0 & 0 & 0 & 0 & 0 & 0 & 0 & 0 & 0 & 0 & 0 & x & 0 & 0 \\ 15 & 0 & 0 & 0 & 0 & 0 & 0 & 0 & 0 & 0 & 0 & 0 & 0 & x & 0 & 0 \end{array}\right] \quad \text{(A-1)}$$

where x denotes different distances between two points. The distance matrix becomes a binary connectivity matrix, when all x are set to 1 (Figure 1a).

$$PPCM = \left[\begin{array}{cccccccccccccccc} & 1 & 2 & 3 & 4 & 5 & 6 & 7 & 8 & 9 & 10 & 11 & 12 & 13 & 14 & 15 \\ 1 & 0 & 1 & 0 & 0 & 0 & 0 & 0 & 0 & 0 & 0 & 0 & 0 & 0 & 0 & 0 \\ 2 & 0 & 0 & 0 & 1 & 1 & 0 & 0 & 0 & 0 & 0 & 0 & 0 & 0 & 0 & 0 \\ 3 & 0 & 0 & 0 & 1 & 0 & 0 & 0 & 0 & 0 & 0 & 0 & 0 & 0 & 0 & 0 \\ 4 & 0 & 1 & 0 & 0 & 1 & 0 & 0 & 0 & 0 & 0 & 0 & 0 & 0 & 0 & 0 \\ 5 & 0 & 1 & 0 & 1 & 0 & 0 & 0 & 0 & 0 & 0 & 0 & 0 & 0 & 0 & 0 \\ 6 & 0 & 0 & 0 & 0 & 1 & 0 & 0 & 0 & 0 & 0 & 0 & 0 & 0 & 0 & 0 \\ 7 & 0 & 0 & 0 & 0 & 1 & 0 & 0 & 1 & 1 & 0 & 0 & 0 & 0 & 0 & 0 \\ 8 & 0 & 0 & 0 & 0 & 0 & 0 & 1 & 0 & 0 & 0 & 1 & 1 & 0 & 0 & 0 \\ 9 & 0 & 0 & 0 & 0 & 0 & 0 & 1 & 0 & 0 & 0 & 0 & 0 & 0 & 0 & 0 \\ 10 & 0 & 0 & 0 & 0 & 0 & 0 & 0 & 0 & 0 & 0 & 1 & 0 & 0 & 0 & 0 \\ 11 & 0 & 0 & 0 & 0 & 0 & 0 & 0 & 1 & 0 & 1 & 0 & 1 & 0 & 0 & 0 \\ 12 & 0 & 0 & 0 & 0 & 0 & 0 & 0 & 1 & 0 & 0 & 1 & 0 & 1 & 0 & 0 \\ 13 & 0 & 0 & 0 & 0 & 0 & 0 & 0 & 0 & 0 & 0 & 0 & 1 & 0 & 1 & 1 \\ 14 & 0 & 0 & 0 & 0 & 0 & 0 & 0 & 0 & 0 & 0 & 0 & 0 & 1 & 0 & 0 \\ 15 & 0 & 0 & 0 & 0 & 0 & 0 & 0 & 0 & 0 & 0 & 0 & 0 & 1 & 0 & 0 \end{array}\right] \quad \text{(A-2)}$$

Line-line adjacent matrix (LLAM) is a binary matrix, whose element is set to 1 if corresponding lines are intersected and 0 otherwise (Figure 1b).

$$LLAM = \left[\begin{array}{cccccccc} & a & b & c & d & e & f & g \\ a & 0 & 1 & 1 & 1 & 1 & 1 & 1 \\ b & 1 & 0 & 0 & 0 & 0 & 0 & 1 \\ c & 1 & 0 & 0 & 0 & 0 & 0 & 0 \\ d & 1 & 0 & 0 & 0 & 1 & 0 & 0 \\ e & 1 & 0 & 0 & 1 & 0 & 0 & 0 \\ f & 1 & 0 & 0 & 0 & 0 & 0 & 0 \\ g & 1 & 1 & 0 & 0 & 0 & 0 & 0 \end{array}\right] \quad \text{(A-3)}$$

Point-point adjacent matrix (PPAM) is a binary matrix indicating whether or not a pair of points share a line in common: 1 if yes and 0 otherwise (Figure 1b).

$$
PPAM = \begin{bmatrix}
 & 1 & 2 & 3 & 4 & 5 & 6 & 7 & 8 & 9 & 10 & 11 & 12 & 13 & 14 & 15 \\
1 & 0 & 1 & 0 & 0 & 1 & 0 & 1 & 1 & 0 & 0 & 0 & 1 & 1 & 0 & 1 \\
2 & 1 & 0 & 0 & 1 & 1 & 0 & 1 & 1 & 0 & 0 & 0 & 1 & 1 & 0 & 1 \\
3 & 0 & 0 & 0 & 1 & 1 & 1 & 0 & 0 & 0 & 0 & 0 & 0 & 0 & 0 & 0 \\
4 & 0 & 1 & 1 & 0 & 1 & 1 & 0 & 0 & 0 & 0 & 0 & 0 & 0 & 0 & 0 \\
5 & 1 & 1 & 1 & 1 & 0 & 1 & 1 & 1 & 0 & 0 & 0 & 1 & 1 & 0 & 1 \\
6 & 0 & 0 & 1 & 1 & 1 & 0 & 0 & 0 & 0 & 0 & 0 & 0 & 0 & 0 & 0 \\
7 & 1 & 1 & 0 & 0 & 1 & 0 & 0 & 1 & 1 & 0 & 0 & 1 & 1 & 0 & 1 \\
8 & 1 & 1 & 0 & 0 & 1 & 0 & 1 & 0 & 0 & 0 & 1 & 1 & 1 & 0 & 1 \\
9 & 0 & 0 & 0 & 0 & 0 & 0 & 1 & 0 & 0 & 0 & 0 & 0 & 0 & 0 & 0 \\
10 & 0 & 0 & 0 & 0 & 0 & 0 & 0 & 0 & 0 & 0 & 1 & 1 & 0 & 0 & 0 \\
11 & 0 & 0 & 0 & 0 & 0 & 0 & 0 & 1 & 0 & 1 & 0 & 1 & 0 & 0 & 0 \\
12 & 1 & 1 & 0 & 0 & 1 & 0 & 1 & 1 & 0 & 1 & 1 & 0 & 1 & 0 & 1 \\
13 & 1 & 1 & 0 & 0 & 1 & 0 & 1 & 1 & 0 & 0 & 0 & 1 & 0 & 1 & 1 \\
14 & 0 & 0 & 0 & 0 & 0 & 0 & 0 & 0 & 0 & 0 & 0 & 0 & 1 & 0 & 0 \\
15 & 1 & 1 & 0 & 0 & 1 & 0 & 1 & 1 & 0 & 0 & 0 & 1 & 1 & 0 & 0
\end{bmatrix} \qquad \text{(A-4)}
$$

Generally, a Line-Point Incidence Matrix (LPIM) shows the relationship between lines and points, i.e., whether or not a point on a line (Figure 1b).

$$
LPIM = \begin{bmatrix}
 & 1 & 2 & 3 & 4 & 5 & 6 & 7 & 8 & 9 & 10 & 11 & 12 & 13 & 14 & 15 \\
a & 1 & 1 & 0 & 0 & 1 & 0 & 1 & 1 & 0 & 0 & 0 & 1 & 1 & 0 & 1 \\
b & 0 & 0 & 1 & 1 & 1 & 1 & 0 & 0 & 0 & 0 & 0 & 0 & 0 & 0 & 0 \\
c & 0 & 0 & 0 & 0 & 0 & 0 & 1 & 0 & 1 & 0 & 0 & 0 & 0 & 0 & 0 \\
d & 0 & 0 & 0 & 0 & 0 & 0 & 0 & 1 & 0 & 0 & 1 & 0 & 0 & 0 & 0 \\
e & 0 & 0 & 0 & 0 & 0 & 0 & 0 & 0 & 0 & 1 & 1 & 1 & 0 & 0 & 0 \\
f & 0 & 0 & 0 & 0 & 0 & 0 & 0 & 0 & 0 & 0 & 0 & 0 & 1 & 1 & 0 \\
g & 0 & 1 & 0 & 1 & 0 & 0 & 0 & 0 & 0 & 0 & 0 & 0 & 0 & 0 & 0
\end{bmatrix} \qquad \text{(A-5)}
$$

The above two adjacency matrices can be easily derived from LPIM using the following operations:

$$LLAM = LPIM * LPIM^{T}$$
$$PPAM = LPIM^{T} * LPIM$$

## Appendix B: Algorithms for forming natural roads with different join principles

```
Input: Segment-based network shape file
Output: road-based network shape file

Sub Main () //main function
  Get all the segments in segment-based network shape file;
  Get randomly a segment as a starting search segment from all the segments;
  While (there is no such a starting search segment) do
     If (this segment is not processed) then
         Change the status of that segment to be processed;
         Get a new segment by calling function 'SearchSegmentByPnt_EveryBestFit' with the old
         segment and it's 'from' direction as parameters;

         /* to call different functions */
         // Get a new segment by calling function 'SearchSegmentByPnt_SelfBestFit' with the old
         // segment and it's 'from' direction as parameters;

         // Get a new segment by calling function 'SearchSegmentByPnt_SelfFit' with the old
         // segment and it's 'from' direction as parameters;

         Get another new segment by calling function 'SearchSegmentByPnt_EveryBestFit' with the
         last new constructed segment and it's 'to' direction as parameters;

         /* to call different functions */
         // Get another new segment by calling function 'SearchSegmentByPnt_SelfBestFit' with
         // the last new constructed segment and it's 'to' direction as parameters;

         // Get another new segment by calling function 'SearchSegmentByPnt_SelfFit' with the
         // last new constructed segment and it's 'to' direction as parameters;

         Create a road with final constructed segment and add to the road-based network shape
         file;
     End
     Get randomly a segment as a starting search segment from all the segments except the
     processed segments;
  End while
End sub
```

## Algorithm I based on the principle of every-best-fit

```
Function SearchSegmentByPnt_EveryBestFit (old segment, direction) as new segment
//a recursive function
  If (direction is 'from' direction) then
     Search point = the from point of old segment;
  Else if (direction is 'to' direction) then
     Search point = the to point of old segment;
  End if
  Use a spatial filter to search for the segments intersected with search point except old
  segment;
  If (there are no intersected segments) then
      Return new segment = old segment;
  Else
      If (the searched segments are all processed) then
         Return new segment = old segment;
      Else
         Exclude the processed segments from the searched segments to get a remained set;
         Calculate the deflection angles (a1) between old segment and every segment in the
         remained set;
         Calculate the deflection angle (a2) of every pair in the remained set;
         Select the segments which meet with the condition a1 < a2;
       // the principle of every-best-fit
         If (There are no selected segments) then
             Return new segment = old segment;
         Else
             Get the minimum deflection angle from a1 and its corresponding segment;
             If (the minimum deflection angle < threshold) then
                Join old segment and that segment into new segment at search point;
                Change the status of that segment to be processed;
                Call function 'SearchSegmentByPnt_EveryBestFit' recursively with new segment
                 and direction as parameters;
            Else
                Return new segment = old segment;
            End If
```

```
        End If
      End If
  End If
End function
```

### Algorithm II based on the principle of self-best-fit

```
Function SearchSegmentByPnt_SelfBestFit (old segment, direction) as new segment
//a recursive function
  If (direction is ‘from’ direction) then
     Search point = the from point of old segment;
  Else if (direction is ‘to’ direction) then
     Search point = the to point of old segment;
  End if
  Use a spatial filter to search for the segments intersected with search point except old
  segment;
  If (there is no intersected segments) then
      Return new segment = old segment;
  Else
      If (the searched segments are all processed) then
         Return new segment = old segment;
      Else
         Exclude the processed segments from the searched segments to get a remained set;
         Calculate the deflection angles between old segment and every segment in the remained
          set;
         Get the minimum deflection angle and its corresponding segment;
         // the principle of Self-best-fit
         If (the minimums deflection angle < threshold) then
             Join old segment and that segment into new segment at search point;
             Change the status of that segment to be processed;
             Call function ‘SearchSegmentByPnt_SelfBestFit’ recursively with new segment and
             direction as parameters;
         Else
           Return new segment = old segment;
         End If
      End If
  End If
End function
```

### Algorithm Ⅲ based on the principle of self-fit

```
Function SearchSegmentByPnt_SelfFit (old segment, direction) as new segment
//a recursive function
  If (direction is ‘from’ direction) then
     Search point = the from point of old segment;
  Else if (direction is ‘to’ direction) then
     Search point = the to point of old segment;
  End if
  Use a spatial filter to search for the segments intersected with search point except old
  segment;
  If (there are no intersected segments) then
      Return new segment = old segment;
  Else
      If (the searched segments are all processed) then
         Return new segment = old segment;
      Else
         Exclude the processed segments from the searched segments to get a remained set;
         Calculate the deflection angles (a1) between old segment and every segment in the
         remained set;
         Select the segments that meet with the condition a1 < threshold
         If (there are not selected segments) then
           Return new segment = old segment;
         Else
           Get randomly a segment from the selected segments;
           // the principle of self-fit
           Join old segment and that segment into new segment at search point;
            Change the status of that segment to be processed;
           Call function ‘SearchSegmentByPnt_SelfFit’ recursively with new segment and
             direction as parameters;
         End if
      End If
  End If
```

```
End function
```

## Appendix C: An introduction to ranking metrics used in the paper

To make the paper self contained, we introduce briefly the seven ranking metrics examined in the paper, including connectivity, control, closeness which leads to both local and global integrations, betweenness, PageRank and weighted PageRank (NOTE: some of the metrics are given in a short format such as Connect, LInteg, GInteg, and Between in the plots). The reader may refer to relevant literature for more details, e.g., Jiang (2006) for space syntax metrics originally developed by Hillier and Hanson (1984), Freeman (1979) for centrality metrics, Langville and Meyer (2006) for the PageRank metric, originally developed by Page and Brin (1998), and Xing and Ghorbani (2004) for the weighted PageRank metric. Note that space syntax metrics with the exception of the control metric are based on centrality metrics, although they are named differently. In what follows, we will outline the linkage.

The connectivity metric is de facto degree centrality, which measures the number of roads that interconnect a given road. In the connectivity graph that represents road-road intersection, connectivity is the number of nodes that link a given node. Formally connectivity is defined by:

$$Cnt_i = k \qquad \text{(C-1)}$$

where $k$ is the number of nodes directly linked to the given node $i$.

The control metric of a node is closely related to the connectivity of the directly linked nodes. Formally it is defined by:

$$Ctr_i = \sum_{j=1}^{k} \frac{1}{Cnt_j} \qquad \text{(C-2)}$$

where $k$ is the number of directly linked nodes (or connectivity) of a considered node, and $Cnt_j$ is the connectivity of the $j$th directly linked node.

The closeness metric measures the smallest number of links from a street to all other streets. In the corresponding connectivity graph, it is the shortest distance from a given node to all other nodes. It is defined by:

$$Cls_i = \frac{n-1}{\sum_{k=1}^{n} d(i,j)} \qquad \text{(C-3)}$$

where $d(i,j)$ is the shortest distance between nodes $i$ and $j$.

The closeness metric becomes a sort of local closeness when considering only nodes within a few steps, instead of all the nodes in the connectivity graph. In this sense, the closeness metric given by (C-3) is defined at a global level, thus global closeness metric so to speak. Both local and global closeness metrics are the base for defining local and global integrations, as used in our experiments.

The betweenness centrality measures to what extent a road is between roads. In the connectivity graph, it reflects the intermediary location of a node along indirect relationships linking other nodes. Formally it is defined by:

$$Btw_i = \sum_{j=1}^{n} \sum_{k=1}^{j-1} \frac{p_{ikj}}{p_{ij}} \qquad \text{(C-4)}$$

where $p_{ij}$ is he number of shortest paths from $i$ to $j$, and $p_{ikj}$ is the number of shortest paths from $i$ to $j$ that pass through k, so $\frac{p_{ikj}}{p_{ij}}$ is the proportion of shortest paths from $i$ to $j$ that pass through $k$.

The PageRank metric is initially defined for web graphs (directed in nature) for ranking individual web pages (Page and Brin 1998). The basic idea of PageRank is that a highly ranked node is one that highly ranked nodes point to. It is defined formally as follows:

$$\Pr_i = \frac{1-d}{n} + d \sum_{j \in ON(i)} \frac{\Pr_j}{n_j} \qquad \text{(C-5)}$$

where $n$ is the total number of nodes; $ON(i)$ is the outlink neighbors (i.e., those nodes that point to node $i$); $Pr_i$ and $Pr_j$ are rank scores of node $i$ and $j$, respectively; $n_j$ denotes the number of outlink nodes of node j; $d$ is a damping factor, which is usually set to 0.85 for ranking web pages.

With the above definition of PageRank, the PageRank of a node at any iteration is evenly divided over the nodes to which it links (or outlink nodes). However, the propagation of the PageRank should follow an uneven rule, i.e., the more popular nodes tend to get a higher proportion. This is exactly the basic motivation of the weighted PageRank (Xing and Ghorbani 2004).

The weighted PageRank is defined as follows:

$$Wpr_i = \frac{1-d}{n} + d \sum_{j \in ON(i)} Wpr_j W_j \qquad \text{(C-6)}$$

where weight $W_j$ is added to propagate a PageRank score from one particular node $i$ to its outlink nodes. This is different from equation (C-5), where a PageRank score is evenly divided among its outlink nodes.

The weight $W_j$ represents the relative popularity of node $j$ among its counterparts, and it is defined as follows:

$$W_j = \frac{w_j}{\sum w(k)} \qquad \text{(C-7)}$$

where $k$ is the counterpart nodes of $j$, $w$ is the weight for individual links, indicating their relative popularity based on the percentage of inlinks and outlinks.